\documentclass[twocolumn,traditabstract,longauth]{aa}
\usepackage{graphicx,amsmath,amsfonts,amssymb}
\usepackage{epsf}
\usepackage{url}
\usepackage{natbib}
\bibpunct{(}{)}{;}{a}{}{,} 

\def\setsymbol#1#2{\expandafter\def\csname #1\endcsname{#2}}
\def\getsymbol#1{\csname #1\endcsname}

\def\Planck{{\it Planck\/}}

\def\allearlypapers{\nocite{planck2011-1.1, planck2011-1.3, planck2011-1.4, planck2011-1.5, planck2011-1.6, planck2011-1.7, planck2011-1.10, planck2011-1.10sup, planck2011-5.1a, planck2011-5.1b, planck2011-5.2a, planck2011-5.2b, planck2011-5.2c, planck2011-6.1, planck2011-6.2, planck2011-6.3a, planck2011-6.4a, planck2011-6.4b, planck2011-6.6, planck2011-7.0, planck2011-7.2, planck2011-7.3, planck2011-7.7a, planck2011-7.7b, planck2011-7.12, planck2011-7.13}}

\newbox\tablebox    \newdimen\tablewidth
\def\leaderfil{\leaders\hbox to 5pt{\hss.\hss}\hfil}
\def\endPlancktable{\tablewidth=\columnwidth 
    $$\hss\copy\tablebox\hss$$
    \vskip-\lastskip\vskip -2pt}

\def\tablenote#1 #2\par{\begingroup \parindent=0.8em
    \abovedisplayshortskip=0pt\belowdisplayshortskip=0pt
    \noindent
    $$\hss\vbox{\hsize\tablewidth \hangindent=\parindent \hangafter=1 \noindent
    \hbox to \parindent{\sup{\rm #1}\hss}\strut#2\strut\par}\hss$$
    \endgroup}
\def\doubleline{\vskip 3pt\hrule \vskip 1.5pt \hrule \vskip 5pt}

\def\L2{\ifmmode L_2\else $L_2$\fi}

\def\DeltaT{\ifmmode \Delta T\else $\Delta T$\fi}
\def\deltat{\ifmmode \Delta t\else $\Delta t$\fi}
\def\fknee{\ifmmode f_{\rm knee}\else $f_{\rm knee}$\fi}
\def\Fmax{\ifmmode F_{\rm max}\else $F_{\rm max}$\fi}
\def\solar{\ifmmode{\rm M}_{\mathord\odot}\else${\rm M}_{\mathord\odot}$\fi}

\def\inv{\ifmmode^{-1}\else$^{-1}$\fi}
\def\mo{\ifmmode^{-1}\else$^{-1}$\fi}
\def\sup#1{\ifmmode ^{\rm #1}\else $^{\rm #1}$\fi}
\def\expo#1{\ifmmode \times 10^{#1}\else $\times 10^{#1}$\fi}
\def\,{\thinspace}
\def\lsim{\mathrel{\raise .4ex\hbox{\rlap{$<$}\lower 1.2ex\hbox{$\sim$}}}}
\def\gsim{\mathrel{\raise .4ex\hbox{\rlap{$>$}\lower 1.2ex\hbox{$\sim$}}}}

\def\simprop{\mathrel{\raise .4ex\hbox{\rlap{$\propto$}\lower 1.2ex\hbox{$\sim$}}}}
\def\deg{\ifmmode^\circ\else$^\circ$\fi}
\def\pdeg{\ifmmode $\setbox0=\hbox{$^{\circ}$}\rlap{\hskip.11\wd0 .}$^{\circ}
          \else \setbox0=\hbox{$^{\circ}$}\rlap{\hskip.11\wd0 .}$^{\circ}$\fi}
\def\arcs{\ifmmode {^{\scriptstyle\prime\prime}}
          \else $^{\scriptstyle\prime\prime}$\fi}
\def\arcm{\ifmmode {^{\scriptstyle\prime}}
          \else $^{\scriptstyle\prime}$\fi}
\newdimen\sa  \newdimen\sb
\def\parcs{\sa=.07em \sb=.03em
     \ifmmode \hbox{\rlap{.}}^{\scriptstyle\prime\kern -\sb\prime}\hbox{\kern -\sa}
     \else \rlap{.}$^{\scriptstyle\prime\kern -\sb\prime}$\kern -\sa\fi}
\def\parcm{\sa=.08em \sb=.03em
     \ifmmode \hbox{\rlap{.}\kern\sa}^{\scriptstyle\prime}\hbox{\kern-\sb}
     \else \rlap{.}\kern\sa$^{\scriptstyle\prime}$\kern-\sb\fi}
\def\ra[#1 #2 #3.#4]{#1\sup{h}#2\sup{m}#3\sup{s}\llap.#4}
\def\dec[#1 #2 #3.#4]{#1\deg#2\arcm#3\arcs\llap.#4}
\def\deco[#1 #2 #3]{#1\deg#2\arcm#3\arcs}
\def\rra[#1 #2]{#1\sup{h}#2\sup{m}}

\def\dots{\relax\ifmmode \ldots\else $\ldots$\fi}
\def\WHzsr{\ifmmode $W\,Hz\mo\,sr\mo$\else W\,Hz\mo\,sr\mo\fi}
\def\mHz{\ifmmode $\,mHz$\else \,mHz\fi}
\def\GHz{\ifmmode $\,GHz$\else \,GHz\fi}
\def\mKs{\ifmmode $\,mK\,s$^{1/2}\else \,mK\,s$^{1/2}$\fi}
\def\muKs{\ifmmode \,\mu$K\,s$^{1/2}\else \,$\mu$K\,s$^{1/2}$\fi}
\def\muKRJs{\ifmmode \,\mu$K$_{\rm RJ}$\,s$^{1/2}\else \,$\mu$K$_{\rm RJ}$\,s$^{1/2}$\fi}
\def\muKHz{\ifmmode \,\mu$K\,Hz$^{-1/2}\else \,$\mu$K\,Hz$^{-1/2}$\fi}
\def\MJysr{\ifmmode \,$MJy\,sr\mo$\else \,MJy\,sr\mo\fi}
\def\MJysrmK{\ifmmode \,$MJy\,sr\mo$\,mK$_{\rm CMB}\mo\else \,MJy\,sr\mo\,mK$_{\rm CMB}\mo$\fi}
\def\microns{\ifmmode \,\mu$m$\else \,$\mu$m\fi}

\def\muK{\ifmmode \,\mu$K$\else \,$\mu$\hbox{K}\fi}
\def\microK{\ifmmode \,\mu$K$\else \,$\mu$\hbox{K}\fi}
\def\muW{\ifmmode \,\mu$W$\else \,$\mu$\hbox{W}\fi}
\def\kms{\ifmmode $\,km\,s$^{-1}\else \,km\,s$^{-1}$\fi}
\def\kmsMpc{\ifmmode $\,\kms\,Mpc\mo$\else \,\kms\,Mpc\mo\fi}
\setsymbol{LFI:center:frequency:70GHz:units}{70.3\,GHz}
\setsymbol{LFI:center:frequency:44GHz:units}{44.1\,GHz}
\setsymbol{LFI:center:frequency:30GHz:units}{28.5\,GHz}

\setsymbol{LFI:center:frequency:70GHz}{70.3}
\setsymbol{LFI:center:frequency:44GHz}{44.1}
\setsymbol{LFI:center:frequency:30GHz}{28.5}

\setsymbol{LFI:center:frequency:LFI18:Rad:M:units}{71.7\GHz}
\setsymbol{LFI:center:frequency:LFI19:Rad:M:units}{67.5\GHz}
\setsymbol{LFI:center:frequency:LFI20:Rad:M:units}{69.2\GHz}
\setsymbol{LFI:center:frequency:LFI21:Rad:M:units}{70.4\GHz}
\setsymbol{LFI:center:frequency:LFI22:Rad:M:units}{71.5\GHz}
\setsymbol{LFI:center:frequency:LFI23:Rad:M:units}{70.8\GHz}
\setsymbol{LFI:center:frequency:LFI24:Rad:M:units}{44.4\GHz}
\setsymbol{LFI:center:frequency:LFI25:Rad:M:units}{44.0\GHz}
\setsymbol{LFI:center:frequency:LFI26:Rad:M:units}{43.9\GHz}
\setsymbol{LFI:center:frequency:LFI27:Rad:M:units}{28.3\GHz}
\setsymbol{LFI:center:frequency:LFI28:Rad:M:units}{28.8\GHz}
\setsymbol{LFI:center:frequency:LFI18:Rad:S:units}{70.1\GHz}
\setsymbol{LFI:center:frequency:LFI19:Rad:S:units}{69.6\GHz}
\setsymbol{LFI:center:frequency:LFI20:Rad:S:units}{69.5\GHz}
\setsymbol{LFI:center:frequency:LFI21:Rad:S:units}{69.5\GHz}
\setsymbol{LFI:center:frequency:LFI22:Rad:S:units}{72.8\GHz}
\setsymbol{LFI:center:frequency:LFI23:Rad:S:units}{71.3\GHz}
\setsymbol{LFI:center:frequency:LFI24:Rad:S:units}{44.1\GHz}
\setsymbol{LFI:center:frequency:LFI25:Rad:S:units}{44.1\GHz}
\setsymbol{LFI:center:frequency:LFI26:Rad:S:units}{44.1\GHz}
\setsymbol{LFI:center:frequency:LFI27:Rad:S:units}{28.5\GHz}
\setsymbol{LFI:center:frequency:LFI28:Rad:S:units}{28.2\GHz}

\setsymbol{LFI:center:frequency:LFI18:Rad:M}{71.7}
\setsymbol{LFI:center:frequency:LFI19:Rad:M}{67.5}
\setsymbol{LFI:center:frequency:LFI20:Rad:M}{69.2}
\setsymbol{LFI:center:frequency:LFI21:Rad:M}{70.4}
\setsymbol{LFI:center:frequency:LFI22:Rad:M}{71.5}
\setsymbol{LFI:center:frequency:LFI23:Rad:M}{70.8}
\setsymbol{LFI:center:frequency:LFI24:Rad:M}{44.4}
\setsymbol{LFI:center:frequency:LFI25:Rad:M}{44.0}
\setsymbol{LFI:center:frequency:LFI26:Rad:M}{43.9}
\setsymbol{LFI:center:frequency:LFI27:Rad:M}{28.3}
\setsymbol{LFI:center:frequency:LFI28:Rad:M}{28.8}
\setsymbol{LFI:center:frequency:LFI18:Rad:S}{70.1}
\setsymbol{LFI:center:frequency:LFI19:Rad:S}{69.6}
\setsymbol{LFI:center:frequency:LFI20:Rad:S}{69.5}
\setsymbol{LFI:center:frequency:LFI21:Rad:S}{69.5}
\setsymbol{LFI:center:frequency:LFI22:Rad:S}{72.8}
\setsymbol{LFI:center:frequency:LFI23:Rad:S}{71.3}
\setsymbol{LFI:center:frequency:LFI24:Rad:S}{44.1}
\setsymbol{LFI:center:frequency:LFI25:Rad:S}{44.1}
\setsymbol{LFI:center:frequency:LFI26:Rad:S}{44.1}
\setsymbol{LFI:center:frequency:LFI27:Rad:S}{28.5}
\setsymbol{LFI:center:frequency:LFI28:Rad:S}{28.2}

\setsymbol{LFI:white:noise:sensitivity:70GHz:units}{134.7\muKs}
\setsymbol{LFI:white:noise:sensitivity:44GHz:units}{164.7\muKs}
\setsymbol{LFI:white:noise:sensitivity:30GHz:units}{143.4\muKs}

\setsymbol{LFI:white:noise:sensitivity:70GHz}{134.7}
\setsymbol{LFI:white:noise:sensitivity:44GHz}{164.7}
\setsymbol{LFI:white:noise:sensitivity:30GHz}{143.4}

\setsymbol{LFI:white:noise:sensitivity:LFI18:Rad:M:units}{512.0\muKs}
\setsymbol{LFI:white:noise:sensitivity:LFI19:Rad:M:units}{581.4\muKs}
\setsymbol{LFI:white:noise:sensitivity:LFI20:Rad:M:units}{590.8\muKs}
\setsymbol{LFI:white:noise:sensitivity:LFI21:Rad:M:units}{455.2\muKs}
\setsymbol{LFI:white:noise:sensitivity:LFI22:Rad:M:units}{492.0\muKs}
\setsymbol{LFI:white:noise:sensitivity:LFI23:Rad:M:units}{507.7\muKs}
\setsymbol{LFI:white:noise:sensitivity:LFI24:Rad:M:units}{462.2\muKs}
\setsymbol{LFI:white:noise:sensitivity:LFI25:Rad:M:units}{413.6\muKs}
\setsymbol{LFI:white:noise:sensitivity:LFI26:Rad:M:units}{478.6\muKs}
\setsymbol{LFI:white:noise:sensitivity:LFI27:Rad:M:units}{277.7\muKs}
\setsymbol{LFI:white:noise:sensitivity:LFI28:Rad:M:units}{312.3\muKs}
\setsymbol{LFI:white:noise:sensitivity:LFI18:Rad:S:units}{465.7\muKs}
\setsymbol{LFI:white:noise:sensitivity:LFI19:Rad:S:units}{555.6\muKs}
\setsymbol{LFI:white:noise:sensitivity:LFI20:Rad:S:units}{623.2\muKs}
\setsymbol{LFI:white:noise:sensitivity:LFI21:Rad:S:units}{564.1\muKs}
\setsymbol{LFI:white:noise:sensitivity:LFI22:Rad:S:units}{534.4\muKs}
\setsymbol{LFI:white:noise:sensitivity:LFI23:Rad:S:units}{542.4\muKs}
\setsymbol{LFI:white:noise:sensitivity:LFI24:Rad:S:units}{399.2\muKs}
\setsymbol{LFI:white:noise:sensitivity:LFI25:Rad:S:units}{392.6\muKs}
\setsymbol{LFI:white:noise:sensitivity:LFI26:Rad:S:units}{418.6\muKs}
\setsymbol{LFI:white:noise:sensitivity:LFI27:Rad:S:units}{302.9\muKs}
\setsymbol{LFI:white:noise:sensitivity:LFI28:Rad:S:units}{285.3\muKs}

\setsymbol{LFI:white:noise:sensitivity:LFI18:Rad:M}{512.0}
\setsymbol{LFI:white:noise:sensitivity:LFI19:Rad:M}{581.4}
\setsymbol{LFI:white:noise:sensitivity:LFI20:Rad:M}{590.8}
\setsymbol{LFI:white:noise:sensitivity:LFI21:Rad:M}{455.2}
\setsymbol{LFI:white:noise:sensitivity:LFI22:Rad:M}{492.0}
\setsymbol{LFI:white:noise:sensitivity:LFI23:Rad:M}{507.7}
\setsymbol{LFI:white:noise:sensitivity:LFI24:Rad:M}{462.2}
\setsymbol{LFI:white:noise:sensitivity:LFI25:Rad:M}{413.6}
\setsymbol{LFI:white:noise:sensitivity:LFI26:Rad:M}{478.6}
\setsymbol{LFI:white:noise:sensitivity:LFI27:Rad:M}{277.7}
\setsymbol{LFI:white:noise:sensitivity:LFI28:Rad:M}{312.3}
\setsymbol{LFI:white:noise:sensitivity:LFI18:Rad:S}{465.7}
\setsymbol{LFI:white:noise:sensitivity:LFI19:Rad:S}{555.6}
\setsymbol{LFI:white:noise:sensitivity:LFI20:Rad:S}{623.2}
\setsymbol{LFI:white:noise:sensitivity:LFI21:Rad:S}{564.1}
\setsymbol{LFI:white:noise:sensitivity:LFI22:Rad:S}{534.4}
\setsymbol{LFI:white:noise:sensitivity:LFI23:Rad:S}{542.4}
\setsymbol{LFI:white:noise:sensitivity:LFI24:Rad:S}{399.2}
\setsymbol{LFI:white:noise:sensitivity:LFI25:Rad:S}{392.6}
\setsymbol{LFI:white:noise:sensitivity:LFI26:Rad:S}{418.6}
\setsymbol{LFI:white:noise:sensitivity:LFI27:Rad:S}{302.9}
\setsymbol{LFI:white:noise:sensitivity:LFI28:Rad:S}{285.3}

\setsymbol{LFI:knee:frequency:70GHz:units}{29.5\mHz}
\setsymbol{LFI:knee:frequency:44GHz:units}{56.2\mHz}
\setsymbol{LFI:knee:frequency:30GHz:units}{113.7\mHz}

\setsymbol{LFI:knee:frequency:70GHz}{29.5}
\setsymbol{LFI:knee:frequency:44GHz}{56.2}
\setsymbol{LFI:knee:frequency:30GHz}{113.7}

\setsymbol{LFI:knee:frequency:LFI18:Rad:M:units}{16.3\mHz}
\setsymbol{LFI:knee:frequency:LFI19:Rad:M:units}{15.1\mHz}
\setsymbol{LFI:knee:frequency:LFI20:Rad:M:units}{18.7\mHz}
\setsymbol{LFI:knee:frequency:LFI21:Rad:M:units}{37.2\mHz}
\setsymbol{LFI:knee:frequency:LFI22:Rad:M:units}{12.7\mHz}
\setsymbol{LFI:knee:frequency:LFI23:Rad:M:units}{34.6\mHz}
\setsymbol{LFI:knee:frequency:LFI24:Rad:M:units}{46.2\mHz}
\setsymbol{LFI:knee:frequency:LFI25:Rad:M:units}{24.9\mHz}
\setsymbol{LFI:knee:frequency:LFI26:Rad:M:units}{67.6\mHz}
\setsymbol{LFI:knee:frequency:LFI27:Rad:M:units}{187.4\mHz}
\setsymbol{LFI:knee:frequency:LFI28:Rad:M:units}{122.2\mHz}
\setsymbol{LFI:knee:frequency:LFI18:Rad:S:units}{17.7\mHz}
\setsymbol{LFI:knee:frequency:LFI19:Rad:S:units}{22.0\mHz}
\setsymbol{LFI:knee:frequency:LFI20:Rad:S:units}{8.7\mHz}
\setsymbol{LFI:knee:frequency:LFI21:Rad:S:units}{25.9\mHz}
\setsymbol{LFI:knee:frequency:LFI22:Rad:S:units}{15.8\mHz}
\setsymbol{LFI:knee:frequency:LFI23:Rad:S:units}{129.8\mHz}
\setsymbol{LFI:knee:frequency:LFI24:Rad:S:units}{100.9\mHz}
\setsymbol{LFI:knee:frequency:LFI25:Rad:S:units}{38.9\mHz}
\setsymbol{LFI:knee:frequency:LFI26:Rad:S:units}{58.9\mHz}
\setsymbol{LFI:knee:frequency:LFI27:Rad:S:units}{104.4\mHz}
\setsymbol{LFI:knee:frequency:LFI28:Rad:S:units}{40.7\mHz}

\setsymbol{LFI:knee:frequency:LFI18:Rad:M}{16.3}
\setsymbol{LFI:knee:frequency:LFI19:Rad:M}{15.1}
\setsymbol{LFI:knee:frequency:LFI20:Rad:M}{18.7}
\setsymbol{LFI:knee:frequency:LFI21:Rad:M}{37.2}
\setsymbol{LFI:knee:frequency:LFI22:Rad:M}{12.7}
\setsymbol{LFI:knee:frequency:LFI23:Rad:M}{34.6}
\setsymbol{LFI:knee:frequency:LFI24:Rad:M}{46.2}
\setsymbol{LFI:knee:frequency:LFI25:Rad:M}{24.9}
\setsymbol{LFI:knee:frequency:LFI26:Rad:M}{67.6}
\setsymbol{LFI:knee:frequency:LFI27:Rad:M}{187.4}
\setsymbol{LFI:knee:frequency:LFI28:Rad:M}{122.2}
\setsymbol{LFI:knee:frequency:LFI18:Rad:S}{17.7}
\setsymbol{LFI:knee:frequency:LFI19:Rad:S}{22.0}
\setsymbol{LFI:knee:frequency:LFI20:Rad:S}{8.7}
\setsymbol{LFI:knee:frequency:LFI21:Rad:S}{25.9}
\setsymbol{LFI:knee:frequency:LFI22:Rad:S}{15.8}
\setsymbol{LFI:knee:frequency:LFI23:Rad:S}{129.8}
\setsymbol{LFI:knee:frequency:LFI24:Rad:S}{100.9}
\setsymbol{LFI:knee:frequency:LFI25:Rad:S}{38.9}
\setsymbol{LFI:knee:frequency:LFI26:Rad:S}{58.9}
\setsymbol{LFI:knee:frequency:LFI27:Rad:S}{104.4}
\setsymbol{LFI:knee:frequency:LFI28:Rad:S}{40.7}

\setsymbol{LFI:slope:70GHz:units}{$-1.03$\mHz}
\setsymbol{LFI:slope:44GHz:units}{$-0.89$\mHz}
\setsymbol{LFI:slope:30GHz:units}{$-0.87$\mHz}

\setsymbol{LFI:slope:70GHz}{$-1.03$}
\setsymbol{LFI:slope:44GHz}{$-0.89$}
\setsymbol{LFI:slope:30GHz}{$-0.87$}

\setsymbol{LFI:slope:LFI18:Rad:M:units}{$-1.04$\mHz}
\setsymbol{LFI:slope:LFI19:Rad:M:units}{$-1.09$\mHz}
\setsymbol{LFI:slope:LFI20:Rad:M:units}{$-0.69$\mHz}
\setsymbol{LFI:slope:LFI21:Rad:M:units}{$-1.56$\mHz}
\setsymbol{LFI:slope:LFI22:Rad:M:units}{$-1.01$\mHz}
\setsymbol{LFI:slope:LFI23:Rad:M:units}{$-0.96$\mHz}
\setsymbol{LFI:slope:LFI24:Rad:M:units}{$-0.83$\mHz}
\setsymbol{LFI:slope:LFI25:Rad:M:units}{$-0.91$\mHz}
\setsymbol{LFI:slope:LFI26:Rad:M:units}{$-0.95$\mHz}
\setsymbol{LFI:slope:LFI27:Rad:M:units}{$-0.87$\mHz}
\setsymbol{LFI:slope:LFI28:Rad:M:units}{$-0.88$\mHz}
\setsymbol{LFI:slope:LFI18:Rad:S:units}{$-1.15$\mHz}
\setsymbol{LFI:slope:LFI19:Rad:S:units}{$-1.00$\mHz}
\setsymbol{LFI:slope:LFI20:Rad:S:units}{$-0.95$\mHz}
\setsymbol{LFI:slope:LFI21:Rad:S:units}{$-0.92$\mHz}
\setsymbol{LFI:slope:LFI22:Rad:S:units}{$-1.01$\mHz}
\setsymbol{LFI:slope:LFI23:Rad:S:units}{$-0.95$\mHz}
\setsymbol{LFI:slope:LFI24:Rad:S:units}{$-0.73$\mHz}
\setsymbol{LFI:slope:LFI25:Rad:S:units}{$-1.16$\mHz}
\setsymbol{LFI:slope:LFI26:Rad:S:units}{$-0.79$\mHz}
\setsymbol{LFI:slope:LFI27:Rad:S:units}{$-0.82$\mHz}
\setsymbol{LFI:slope:LFI28:Rad:S:units}{$-0.91$\mHz}

\setsymbol{LFI:slope:LFI18:Rad:M}{$-1.04$}
\setsymbol{LFI:slope:LFI19:Rad:M}{$-1.09$}
\setsymbol{LFI:slope:LFI20:Rad:M}{$-0.69$}
\setsymbol{LFI:slope:LFI21:Rad:M}{$-1.56$}
\setsymbol{LFI:slope:LFI22:Rad:M}{$-1.01$}
\setsymbol{LFI:slope:LFI23:Rad:M}{$-0.96$}
\setsymbol{LFI:slope:LFI24:Rad:M}{$-0.83$}
\setsymbol{LFI:slope:LFI25:Rad:M}{$-0.91$}
\setsymbol{LFI:slope:LFI26:Rad:M}{$-0.95$}
\setsymbol{LFI:slope:LFI27:Rad:M}{$-0.87$}
\setsymbol{LFI:slope:LFI28:Rad:M}{$-0.88$}
\setsymbol{LFI:slope:LFI18:Rad:S}{$-1.15$}
\setsymbol{LFI:slope:LFI19:Rad:S}{$-1.00$}
\setsymbol{LFI:slope:LFI20:Rad:S}{$-0.95$}
\setsymbol{LFI:slope:LFI21:Rad:S}{$-0.92$}
\setsymbol{LFI:slope:LFI22:Rad:S}{$-1.01$}
\setsymbol{LFI:slope:LFI23:Rad:S}{$-0.95$}
\setsymbol{LFI:slope:LFI24:Rad:S}{$-0.73$}
\setsymbol{LFI:slope:LFI25:Rad:S}{$-1.16$}
\setsymbol{LFI:slope:LFI26:Rad:S}{$-0.79$}
\setsymbol{LFI:slope:LFI27:Rad:S}{$-0.82$}
\setsymbol{LFI:slope:LFI28:Rad:S}{$-0.91$}

\setsymbol{LFI:FWHM:70GHz:units}{13\parcm01}
\setsymbol{LFI:FWHM:44GHz:units}{27\parcm92}
\setsymbol{LFI:FWHM:30GHz:units}{32\parcm65}

\setsymbol{LFI:FWHM:70GHz}{13.01}
\setsymbol{LFI:FWHM:44GHz}{27.92}
\setsymbol{LFI:FWHM:30GHz}{32.65}

\setsymbol{LFI:FWHM:LFI18:units}{13\parcm39}
\setsymbol{LFI:FWHM:LFI19:units}{13\parcm01}
\setsymbol{LFI:FWHM:LFI20:units}{12\parcm75}
\setsymbol{LFI:FWHM:LFI21:units}{12\parcm74}
\setsymbol{LFI:FWHM:LFI22:units}{12\parcm87}
\setsymbol{LFI:FWHM:LFI23:units}{13\parcm27}
\setsymbol{LFI:FWHM:LFI24:units}{22\parcm98}
\setsymbol{LFI:FWHM:LFI25:units}{30\parcm46}
\setsymbol{LFI:FWHM:LFI26:units}{30\parcm31}
\setsymbol{LFI:FWHM:LFI27:units}{32\parcm65}
\setsymbol{LFI:FWHM:LFI28:units}{32\parcm66}

\setsymbol{LFI:FWHM:LFI18}{13.39}
\setsymbol{LFI:FWHM:LFI19}{13.01}
\setsymbol{LFI:FWHM:LFI20}{12.75}
\setsymbol{LFI:FWHM:LFI21}{12.74}
\setsymbol{LFI:FWHM:LFI22}{12.87}
\setsymbol{LFI:FWHM:LFI23}{13.27}
\setsymbol{LFI:FWHM:LFI24}{22.98}
\setsymbol{LFI:FWHM:LFI25}{30.46}
\setsymbol{LFI:FWHM:LFI26}{30.31}
\setsymbol{LFI:FWHM:LFI27}{32.65}
\setsymbol{LFI:FWHM:LFI28}{32.66}

\setsymbol{LFI:FWHM:uncertainty:LFI18:units}{0.170\arcm}
\setsymbol{LFI:FWHM:uncertainty:LFI19:units}{0.174\arcm}
\setsymbol{LFI:FWHM:uncertainty:LFI20:units}{0.170\arcm}
\setsymbol{LFI:FWHM:uncertainty:LFI21:units}{0.156\arcm}
\setsymbol{LFI:FWHM:uncertainty:LFI22:units}{0.164\arcm}
\setsymbol{LFI:FWHM:uncertainty:LFI23:units}{0.171\arcm}
\setsymbol{LFI:FWHM:uncertainty:LFI24:units}{0.652\arcm}
\setsymbol{LFI:FWHM:uncertainty:LFI25:units}{1.075\arcm}
\setsymbol{LFI:FWHM:uncertainty:LFI26:units}{1.131\arcm}
\setsymbol{LFI:FWHM:uncertainty:LFI27:units}{1.266\arcm}
\setsymbol{LFI:FWHM:uncertainty:LFI28:units}{1.287\arcm}

\setsymbol{LFI:FWHM:uncertainty:LFI18}{0.170}
\setsymbol{LFI:FWHM:uncertainty:LFI19}{0.174}
\setsymbol{LFI:FWHM:uncertainty:LFI20}{0.170}
\setsymbol{LFI:FWHM:uncertainty:LFI21}{0.156}
\setsymbol{LFI:FWHM:uncertainty:LFI22}{0.164}
\setsymbol{LFI:FWHM:uncertainty:LFI23}{0.171}
\setsymbol{LFI:FWHM:uncertainty:LFI24}{0.652}
\setsymbol{LFI:FWHM:uncertainty:LFI25}{1.075}
\setsymbol{LFI:FWHM:uncertainty:LFI26}{1.131}
\setsymbol{LFI:FWHM:uncertainty:LFI27}{1.266}
\setsymbol{LFI:FWHM:uncertainty:LFI28}{1.287}

\setsymbol{HFI:center:frequency:100GHz:units}{100\,GHz}
\setsymbol{HFI:center:frequency:143GHz:units}{143\,GHz}
\setsymbol{HFI:center:frequency:217GHz:units}{217\,GHz}
\setsymbol{HFI:center:frequency:353GHz:units}{353\,GHz}
\setsymbol{HFI:center:frequency:545GHz:units}{545\,GHz}
\setsymbol{HFI:center:frequency:857GHz:units}{857\,GHz}

\setsymbol{HFI:center:frequency:100GHz}{100}
\setsymbol{HFI:center:frequency:143GHz}{143}
\setsymbol{HFI:center:frequency:217GHz}{217}
\setsymbol{HFI:center:frequency:353GHz}{353}
\setsymbol{HFI:center:frequency:545GHz}{545}
\setsymbol{HFI:center:frequency:857GHz}{857}

\setsymbol{HFI:Ndetectors:100GHz}{8}
\setsymbol{HFI:Ndetectors:143GHz}{11}
\setsymbol{HFI:Ndetectors:217GHz}{12}
\setsymbol{HFI:Ndetectors:353GHz}{12}
\setsymbol{HFI:Ndetectors:545GHz}{3}
\setsymbol{HFI:Ndetectors:857GHz}{4}

\setsymbol{HFI:FWHM:Maps:100GHz:units}{9\parcm88}
\setsymbol{HFI:FWHM:Maps:143GHz:units}{7\parcm18}
\setsymbol{HFI:FWHM:Maps:217GHz:units}{4\parcm87}
\setsymbol{HFI:FWHM:Maps:353GHz:units}{4\parcm65}
\setsymbol{HFI:FWHM:Maps:545GHz:units}{4\parcm72}
\setsymbol{HFI:FWHM:Maps:857GHz:units}{4\parcm39}
\setsymbol{HFI:FWHM:Maps:100GHz}{9.88}
\setsymbol{HFI:FWHM:Maps:143GHz}{7.18}
\setsymbol{HFI:FWHM:Maps:217GHz}{4.87}
\setsymbol{HFI:FWHM:Maps:353GHz}{4.65}
\setsymbol{HFI:FWHM:Maps:545GHz}{4.72}
\setsymbol{HFI:FWHM:Maps:857GHz}{4.39}

\setsymbol{HFI:beam:ellipticity:Maps:100GHz}{1.15}
\setsymbol{HFI:beam:ellipticity:Maps:143GHz}{1.01}
\setsymbol{HFI:beam:ellipticity:Maps:217GHz}{1.06}
\setsymbol{HFI:beam:ellipticity:Maps:353GHz}{1.05}
\setsymbol{HFI:beam:ellipticity:Maps:545GHz}{1.14}
\setsymbol{HFI:beam:ellipticity:Maps:857GHz}{1.19}

\setsymbol{HFI:FWHM:Mars:100GHz:units}{9\parcm37}
\setsymbol{HFI:FWHM:Mars:143GHz:units}{7\parcm04}
\setsymbol{HFI:FWHM:Mars:217GHz:units}{4\parcm68}
\setsymbol{HFI:FWHM:Mars:353GHz:units}{4\parcm43}
\setsymbol{HFI:FWHM:Mars:545GHz:units}{3\parcm80}
\setsymbol{HFI:FWHM:Mars:857GHz:units}{3\parcm67}

\setsymbol{HFI:FWHM:Mars:100GHz}{9.37}
\setsymbol{HFI:FWHM:Mars:143GHz}{7.04}
\setsymbol{HFI:FWHM:Mars:217GHz}{4.68}
\setsymbol{HFI:FWHM:Mars:353GHz}{4.43}
\setsymbol{HFI:FWHM:Mars:545GHz}{3.80}
\setsymbol{HFI:FWHM:Mars:857GHz}{3.67}

\setsymbol{HFI:beam:ellipticity:Mars:100GHz}{1.18}
\setsymbol{HFI:beam:ellipticity:Mars:143GHz}{1.03}
\setsymbol{HFI:beam:ellipticity:Mars:217GHz}{1.14}
\setsymbol{HFI:beam:ellipticity:Mars:353GHz}{1.09}
\setsymbol{HFI:beam:ellipticity:Mars:545GHz}{1.25}
\setsymbol{HFI:beam:ellipticity:Mars:857GHz}{1.03}

\setsymbol{HFI:CMB:relative:calibration:100GHz}{$\lsim 1\%$}
\setsymbol{HFI:CMB:relative:calibration:143GHz}{$\lsim 1\%$}
\setsymbol{HFI:CMB:relative:calibration:217GHz}{$\lsim 1\%$}
\setsymbol{HFI:CMB:relative:calibration:353GHz}{$\lsim 1\%$}
\setsymbol{HFI:CMB:relative:calibration:545GHz}{}
\setsymbol{HFI:CMB:relative:calibration:857GHz}{}

\setsymbol{HFI:CMB:absolute:calibration:100GHz}{$\lsim 2\%$}
\setsymbol{HFI:CMB:absolute:calibration:143GHz}{$\lsim 2\%$}
\setsymbol{HFI:CMB:absolute:calibration:217GHz}{$\lsim 2\%$}
\setsymbol{HFI:CMB:absolute:calibration:353GHz}{$\lsim 2\%$}
\setsymbol{HFI:CMB:absolute:calibration:545GHz}{}
\setsymbol{HFI:CMB:absolute:calibration:857GHz}{}

\setsymbol{HFI:FIRAS:gain:calibration:accuracy:statistical:100GHz}{}
\setsymbol{HFI:FIRAS:gain:calibration:accuracy:statistical:143GHz}{}
\setsymbol{HFI:FIRAS:gain:calibration:accuracy:statistical:217GHz}{}
\setsymbol{HFI:FIRAS:gain:calibration:accuracy:statistical:353GHz}{2.5\%}
\setsymbol{HFI:FIRAS:gain:calibration:accuracy:statistical:545GHz}{1\%}
\setsymbol{HFI:FIRAS:gain:calibration:accuracy:statistical:857GHz}{0.5\%}

\setsymbol{HFI:FIRAS:gain:calibration:accuracy:systematic:100GHz}{}
\setsymbol{HFI:FIRAS:gain:calibration:accuracy:systematic:143GHz}{}
\setsymbol{HFI:FIRAS:gain:calibration:accuracy:systematic:217GHz}{}
\setsymbol{HFI:FIRAS:gain:calibration:accuracy:systematic:353GHz}{}
\setsymbol{HFI:FIRAS:gain:calibration:accuracy:systematic:545GHz}{7\%}
\setsymbol{HFI:FIRAS:gain:calibration:accuracy:systematic:857GHz}{7\%}

\setsymbol{HFI:FIRAS:zero:point:accuracy:100GHz:units}{0.8\MJysr}
\setsymbol{HFI:FIRAS:zero:point:accuracy:143GHz:units}{}
\setsymbol{HFI:FIRAS:zero:point:accuracy:217GHz:units}{}
\setsymbol{HFI:FIRAS:zero:point:accuracy:353GHz:units}{1.4\MJysr}
\setsymbol{HFI:FIRAS:zero:point:accuracy:545GHz:units}{2.2\MJysr}
\setsymbol{HFI:FIRAS:zero:point:accuracy:857GHz:units}{1.7\MJysr}

\setsymbol{HFI:FIRAS:zero:point:accuracy:100GHz}{0.8}
\setsymbol{HFI:FIRAS:zero:point:accuracy:143GHz}{}
\setsymbol{HFI:FIRAS:zero:point:accuracy:217GHz}{}
\setsymbol{HFI:FIRAS:zero:point:accuracy:353GHz}{1.4}
\setsymbol{HFI:FIRAS:zero:point:accuracy:545GHz}{2.2}
\setsymbol{HFI:FIRAS:zero:point:accuracy:857GHz}{1.7}

\setsymbol{HFI:unit:conversion:100GHz:units}{0.2415\MJysrmK}
\setsymbol{HFI:unit:conversion:143GHz:units}{0.3694\MJysrmK}
\setsymbol{HFI:unit:conversion:217GHz:units}{0.4811\MJysrmK}
\setsymbol{HFI:unit:conversion:353GHz:units}{0.2883\MJysrmK}
\setsymbol{HFI:unit:conversion:545GHz:units}{0.05826\MJysrmK}
\setsymbol{HFI:unit:conversion:857GHz:units}{0.002238\MJysrmK}

\setsymbol{HFI:unit:conversion:100GHz}{0.2415}
\setsymbol{HFI:unit:conversion:143GHz}{0.3694}
\setsymbol{HFI:unit:conversion:217GHz}{0.4811}
\setsymbol{HFI:unit:conversion:353GHz}{0.2883}
\setsymbol{HFI:unit:conversion:545GHz}{0.05826}
\setsymbol{HFI:unit:conversion:857GHz}{0.002238}

\setsymbol{HFI:colour:correction:alpha=-2:V1.01:100GHz}{0.9893}
\setsymbol{HFI:colour:correction:alpha=-2:V1.01:143GHz}{0.9759}
\setsymbol{HFI:colour:correction:alpha=-2:V1.01:217GHz}{1.0007}
\setsymbol{HFI:colour:correction:alpha=-2:V1.01:353GHz}{1.0028}
\setsymbol{HFI:colour:correction:alpha=-2:V1.01:545GHz}{1.0019}
\setsymbol{HFI:colour:correction:alpha=-2:V1.01:857GHz}{0.9889}

\setsymbol{HFI:colour:correction:alpha=0:V1.01:100GHz}{1.0008}
\setsymbol{HFI:colour:correction:alpha=0:V1.01:143GHz}{1.0148}
\setsymbol{HFI:colour:correction:alpha=0:V1.01:217GHz}{0.9909}
\setsymbol{HFI:colour:correction:alpha=0:V1.01:353GHz}{0.9888}
\setsymbol{HFI:colour:correction:alpha=0:V1.01:545GHz}{0.9878}
\setsymbol{HFI:colour:correction:alpha=0:V1.01:857GHz}{1.0014}

\begin{document}
\author{\small
Planck Collaboration:
P.~A.~R.~Ade\inst{75}
\and
N.~Aghanim\inst{51}
\and
F.~Arg\"{u}eso\inst{15}
\and
M.~Arnaud\inst{63}
\and
M.~Ashdown\inst{61, 4}
\and
J.~Aumont\inst{51}
\and
C.~Baccigalupi\inst{73}
\and
A.~Balbi\inst{31}
\and
A.~J.~Banday\inst{79, 8, 68}
\and
R.~B.~Barreiro\inst{57}
% ...
}
\institute{\small
Aalto University Mets\"{a}hovi Radio Observatory, Mets\"{a}hovintie 114, FIN-02540 Kylm\"{a}l\"{a}, Finland\\
\and
Agenzia Spaziale Italiana Science Data Center, c/o ESRIN, via Galileo Galilei, Frascati, Italy\\
\and
Astroparticule et Cosmologie, CNRS (UMR7164), Universit\'{e} Denis Diderot Paris 7, B\^{a}timent Condorcet, 10 rue A. Domon et L\'{e}onie Duquet, Paris, France\\
\and
Astrophysics Group, Cavendish Laboratory, University of Cambridge, J J Thomson Avenue, Cambridge CB3 0HE, U.K.\\
\and
Atacama Large Millimeter/submillimeter Array, ALMA Santiago Central Offices, Alonso de Cordova 3107, Vitacura, Casilla 763 0355, Santiago, Chile\\
\and
Australia Telescope National Facility, CSIRO, P.O. Box 76, Epping, NSW 1710, Australia\\
\and
CITA, University of Toronto, 60 St. George St., Toronto, ON M5S 3H8, Canada\\
\and
CNRS, IRAP, 9 Av. colonel Roche, BP 44346, F-31028 Toulouse cedex 4, France\\
\and
California Institute of Technology, Pasadena, California, U.S.A.\\
\and
Centre of Mathematics for Applications, University of Oslo, Blindern, Oslo, Norway\\
\and
DAMTP, University of Cambridge, Centre for Mathematical Sciences, Wilberforce Road, Cambridge CB3 0WA, U.K.\\
\and
DSM/Irfu/SPP, CEA-Saclay, F-91191 Gif-sur-Yvette Cedex, France\\
\and
DTU Space, National Space Institute, Juliane Mariesvej 30, Copenhagen, Denmark\\
\and
Departamento de F\'{\i}sica, Universidad de Oviedo, Avda. Calvo Sotelo s/n, Oviedo, Spain\\
\and
Departamento de Matem\'{a}ticas, Universidad de Oviedo, Avda. Calvo Sotelo s/n, Oviedo, Spain\\
\and
Department of Astronomy and Astrophysics, University of Toronto, 50 Saint George Street, Toronto, Ontario, Canada\\
\and
Department of Physics \& Astronomy, The Open University, Milton Keynes, MK7 6AA, U.K.\\
\and
Department of Physics \& Astronomy, University of British Columbia, 6224 Agricultural Road, Vancouver, British Columbia, Canada\\
\and
Department of Physics and Astronomy, University of Southern California, Los Angeles, California, U.S.A.\\
\and
Department of Physics, Gustaf H\"{a}llstr\"{o}min katu 2a, University of Helsinki, Helsinki, Finland\\
\and
Department of Physics, Purdue University, 525 Northwestern Avenue, West Lafayette, Indiana, U.S.A.\\
\and
Department of Physics, University of California, Berkeley, California, U.S.A.\\
\and
Department of Physics, University of California, One Shields Avenue, Davis, California, U.S.A.\\
\and
Department of Physics, University of California, Santa Barbara, California, U.S.A.\\
\and
Department of Physics, University of Illinois at Urbana-Champaign, 1110 West Green Street, Urbana, Illinois, U.S.A.\\
\and
Dipartimento di Fisica G. Galilei, Universit\`{a} degli Studi di Padova, via Marzolo 8, 35131 Padova, Italy\\
\and
Dipartimento di Fisica, Universit\`{a} La Sapienza, P. le A. Moro 2, Roma, Italy\\
\and
Dipartimento di Fisica, Universit\`{a} degli Studi di Milano, Via Celoria, 16, Milano, Italy\\
\and
Dipartimento di Fisica, Universit\`{a} degli Studi di Trieste, via A. Valerio 2, Trieste, Italy\\
\and
Dipartimento di Fisica, Universit\`{a} di Ferrara, Via Saragat 1, 44122 Ferrara, Italy\\
\and
Dipartimento di Fisica, Universit\`{a} di Roma Tor Vergata, Via della Ricerca Scientifica, 1, Roma, Italy\\
\and
Discovery Center, Niels Bohr Institute, Blegdamsvej 17, Copenhagen, Denmark\\
\and
Dpto. Astrof\'{i}sica, Universidad de La Laguna (ULL), E-38206 La Laguna, Tenerife, Spain\\
\and
European Southern Observatory, ESO Vitacura, Alonso de Cordova 3107, Vitacura, Casilla 19001, Santiago, Chile\\
\and
European Space Agency, ESAC, Planck Science Office, Camino bajo del Castillo, s/n, Urbanizaci\'{o}n Villafranca del Castillo, Villanueva de la Ca\~{n}ada, Madrid, Spain\\
\and
European Space Agency, ESTEC, Keplerlaan 1, 2201 AZ Noordwijk, The Netherlands\\
\and
Haverford College Astronomy Department, 370 Lancaster Avenue, Haverford, Pennsylvania, U.S.A.\\
\and
Helsinki Institute of Physics, Gustaf H\"{a}llstr\"{o}min katu 2, University of Helsinki, Helsinki, Finland\\
\and
IFSI/INAF, via del Fosso Cavaliere 100, 00133, Roma, Italy\\
\and
INAF - Osservatorio Astrofisico di Catania, Via S. Sofia 78, Catania, Italy\\
\and
INAF - Osservatorio Astronomico di Padova, Vicolo dell'Osservatorio 5, Padova, Italy\\
\and
INAF - Osservatorio Astronomico di Roma, via di Frascati 33, Monte Porzio Catone, Italy\\
\and
INAF - Osservatorio Astronomico di Trieste, Via G.B. Tiepolo 11, Trieste, Italy\\
\and
INAF/IASF Bologna, Via Gobetti 101, Bologna, Italy\\
\and
INAF/IASF Milano, Via E. Bassini 15, Milano, Italy\\
\and
INRIA, Laboratoire de Recherche en Informatique, Universit\'{e} Paris-Sud 11, B\^{a}timent 490, 91405 Orsay Cedex, France\\
\and
IPAG: Institut de Plan\'{e}tologie et d'Astrophysique de Grenoble, Universit\'{e} Joseph Fourier, Grenoble 1 / CNRS-INSU, UMR 5274, Grenoble, F-38041, France\\
\and
ISDC Data Centre for Astrophysics, University of Geneva, ch. d'Ecogia 16, Versoix, Switzerland\\
\and
Imperial College London, Astrophysics group, Blackett Laboratory, Prince Consort Road, London, SW7 2AZ, U.K.\\
\and
Infrared Processing and Analysis Center, California Institute of Technology, Pasadena, CA 91125, U.S.A.\\
\and
Institut d'Astrophysique Spatiale, CNRS (UMR8617) Universit\'{e} Paris-Sud 11, B\^{a}timent 121, Orsay, France\\
\and
Institut d'Astrophysique de Paris, CNRS UMR7095, Universit\'{e} Pierre \& Marie Curie, 98 bis boulevard Arago, Paris, France\\
\and
Institute of Astronomy and Astrophysics, Academia Sinica, Taipei, Taiwan\\
\and
Institute of Astronomy, University of Cambridge, Madingley Road, Cambridge CB3 0HA, U.K.\\
\and
Institute of Theoretical Astrophysics, University of Oslo, Blindern, Oslo, Norway\\
\and
Instituto de Astrof\'{\i}sica de Canarias, C/V\'{\i}a L\'{a}ctea s/n, La Laguna, Tenerife, Spain\\
\and
Instituto de F\'{\i}sica de Cantabria (CSIC-Universidad de Cantabria), Avda. de los Castros s/n, Santander, Spain\\
\and
Istituto di Fisica del Plasma, CNR-ENEA-EURATOM Association, Via R. Cozzi 53, Milano, Italy\\
\and
Jet Propulsion Laboratory, California Institute of Technology, 4800 Oak Grove Drive, Pasadena, California, U.S.A.\\
\and
Jodrell Bank Centre for Astrophysics, Alan Turing Building, School of Physics and Astronomy, The University of Manchester, Oxford Road, Manchester, M13 9PL, U.K.\\
\and
Kavli Institute for Cosmology Cambridge, Madingley Road, Cambridge, CB3 0HA, U.K.\\
\and
LERMA, CNRS, Observatoire de Paris, 61 Avenue de l'Observatoire, Paris, France\\
\and
Laboratoire AIM, IRFU/Service d'Astrophysique - CEA/DSM - CNRS - Universit\'{e} Paris Diderot, B\^{a}t. 709, CEA-Saclay, F-91191 Gif-sur-Yvette Cedex, France\\
\and
Laboratoire Traitement et Communication de l'Information, CNRS (UMR 5141) and T\'{e}l\'{e}com ParisTech, 46 rue Barrault F-75634 Paris Cedex 13, France\\
\and
Laboratoire de Physique Subatomique et de Cosmologie, CNRS, Universit\'{e} Joseph Fourier Grenoble I, 53 rue des Martyrs, Grenoble, France\\
\and
Laboratoire de l'Acc\'{e}l\'{e}rateur Lin\'{e}aire, Universit\'{e} Paris-Sud 11, CNRS/IN2P3, Orsay, France\\
\and
Lawrence Berkeley National Laboratory, Berkeley, California, U.S.A.\\
\and
Max-Planck-Institut f\"{u}r Astrophysik, Karl-Schwarzschild-Str. 1, 85741 Garching, Germany\\
\and
MilliLab, VTT Technical Research Centre of Finland, Tietotie 3, Espoo, Finland\\
\and
National University of Ireland, Department of Experimental Physics, Maynooth, Co. Kildare, Ireland\\
\and
Niels Bohr Institute, Blegdamsvej 17, Copenhagen, Denmark\\
\and
Observational Cosmology, Mail Stop 367-17, California Institute of Technology, Pasadena, CA, 91125, U.S.A.\\
\and
SISSA, Astrophysics Sector, via Bonomea 265, 34136, Trieste, Italy\\
\and
SUPA, Institute for Astronomy, University of Edinburgh, Royal Observatory, Blackford Hill, Edinburgh EH9 3HJ, U.K.\\
\and
School of Physics and Astronomy, Cardiff University, Queens Buildings, The Parade, Cardiff, CF24 3AA, U.K.\\
\and
Space Sciences Laboratory, University of California, Berkeley, California, U.S.A.\\
\and
Spitzer Science Center, 1200 E. California Blvd., Pasadena, California, U.S.A.\\
\and
Stanford University, Dept of Physics, Varian Physics Bldg, 382 Via Pueblo Mall, Stanford, California, U.S.A.\\
\and
Universit\'{e} de Toulouse, UPS-OMP, IRAP, F-31028 Toulouse cedex 4, France\\
% ...
}

  \title{\Planck\ Early Results: Statistical properties of
  extragalactic radio sources in the \Planck\ Early Release Compact Source Catalogue}
\authorrunning{\Planck\ Collaboration}
\titlerunning{Statistical properties of ERS}

   \date{....}

\abstract 
{The data reported in \Planck's Early Release Compact Source
Catalogue (ERCSC) are exploited to measure the number counts
($dN/dS$) of extragalactic radio sources at 30, 44, 70,
100, 143 and 217 GHz. Due to the full-sky nature of the catalogue,
this measurement extends to the rarest and brightest sources in the
sky. At lower frequencies (30, 44, and 70\,GHz) our counts are in
very good agreement with estimates based on \textit{WMAP} data,
being somewhat deeper at 30 and 70\,GHz, and somewhat shallower at
44\,GHz. \Planck's source counts at 143 and 217\,GHz join smoothly
with the fainter ones provided by the SPT  and ACT
 surveys over small fractions of the sky. An analysis of source
spectra, exploiting \Planck's uniquely broad spectral coverage, finds clear evidence
of a steepening of the mean spectral index above about 70\,GHz. This implies that,
at these frequencies, the contamination of the CMB power spectrum by radio sources
below the detection limit is significantly lower than previously estimated.
}


\maketitle
\allearlypapers

\section{Introduction}
\Planck\footnote{\Planck\ (http://www.esa.int/\Planck ) is a project
of the European Space Agency (ESA) with instruments provided by two
scientific consortia funded by ESA member states (in particular the
lead countries: France and Italy) with contributions from NASA
(USA), and telescope reflectors provided in a collaboration between
ESA and a scientific consortium led and funded by Denmark.}
\citep{tauber2010a, planck2011-1.1} is the third-generation space
mission to measure the anisotropy of the cosmic microwave background
(CMB).  It observes the sky in nine frequency bands covering
30--857\,GHz with high sensitivity and angular resolution from
31\arcm\ to 5\arcm.  The Low Frequency Instrument (LFI;
\citealt{Mandolesi2010, Bersanelli2010, planck2011-1.4}) covers the
30, 44, and 70\,GHz bands with amplifiers cooled to 20\,\hbox{K}.
The High Frequency Instrument (HFI; \citealt{Lamarre2010,
planck2011-1.5}) covers the 100, 143, 217, 353, 545, and 857\,GHz
bands with bolometers cooled to 0.1\,\hbox{K}.  Polarization is
measured in all but the highest two bands \citep{Leahy2010,
Rosset2010}.  A combination of radiative cooling and three
mechanical coolers produces the temperatures needed for the
detectors and optics \citep{planck2011-1.3}.  Two data processing
centres (DPCs) check and calibrate the data and make maps of the sky
\citep{planck2011-1.7, planck2011-1.6}.  \Planck's sensitivity,
angular resolution, and frequency coverage make it a powerful
instrument for galactic and extragalactic astrophysics as well as
cosmology.  Early astrophysics results are given in Planck
Collaboration, 2011h--z.

The \Planck\ Early Release Compact Source Catalogue (ERCSC,
\cite{planck2011-1.10}) reports data on sources detected during the
first 1.6 full-sky surveys, and thus offers, among other things, the
opportunity of studying the statistical properties of extragalactic
sources over a broad frequency range never fully explored by blind
surveys. We will focus here on counts of extragalactic radio sources
and on their spectral properties in the 30--217\,GHz
range.\footnote{In all our calculations we have used the effective
central frequencies for the \Planck\ channels
\citep{planck2011-1.4,planck2011-1.5}, although we indicate their
nominal values. The most relevant difference is at 30\,GHz, with a
central frequency of \getsymbol{LFI:center:frequency:30GHz}.}

Although knowledge of the statistical properties at high radio
frequency for this population of extragalactic sources has greatly
improved in the recent past -- thanks to many ground-based
observational campaigns and to the Wilkinson Microwave Anisotropy
Probe (\textit{WMAP}) surveys from space -- above about 70\,GHz these
properties are still largely unknown or very uncertain. This
is essentially due to the fact that very large area surveys at mm
wavelengths are made difficult by the small fields of view of
ground-based radio telescopes and by the long integration times
required.

The most recent estimates on source number counts up to $\sim50-70$
\,GHz, and the optical identifications of the corresponding bright
point sources (see, e.g., \cite{Massardi08,Massardi10}), show that
these counts are dominated by radio sources whose average spectral
index is ``flat'', i.e., $\alpha\simeq 0.0$ (with the usual convention
$S_\nu\propto\nu^\alpha$). This result confirms that the underlying
source population is essentially made of Flat Spectrum Radio Quasars
(FSRQ) and BL Lac objects, collectively called blazars,\footnote{Blazars
are jet-dominated extragalactic objects
characterized by a strongly variable and polarized emission of the
non-thermal radiation, from low radio energies up to the high energy
gamma rays \citep{UrryPadovani95}. Detailed analyses of Spectral
Energy Distributions (SEDs) of complete blazar samples built by
using simultaneous \Planck\ , Swift and Fermi data are given in
\citep{planck2011-6.3a}.} with minor contributions
coming from other source populations \citep{Toffolatti98,DeZotti05}.
At frequencies $> 100$\,GHz, however, there is now new information for
sources with flux densities below about $1\,$Jy coming from the South Pole
Telescope (SPT) collaboration \citep{Vieira10}, with surveys over 87
deg$^2$ at 150 and 220\,GHz, and from the Atacama Cosmology Telescope
(ACT) survey over 455 deg$^2$ at 148\,GHz \citep{Marriage10}.

\begin{figure*}
\begin{center}
\includegraphics[width=0.45\textwidth]{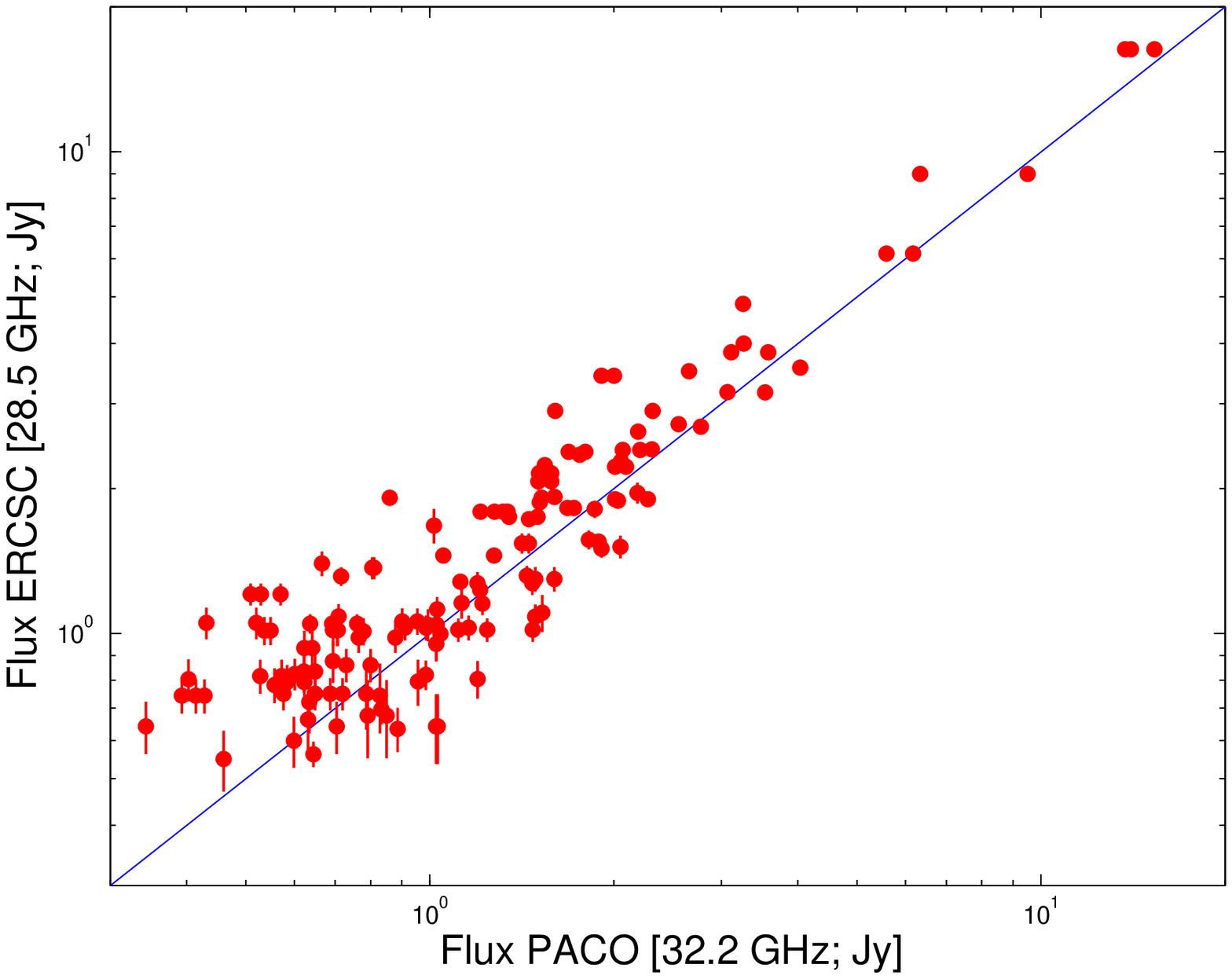}
\includegraphics[width=0.45\textwidth]{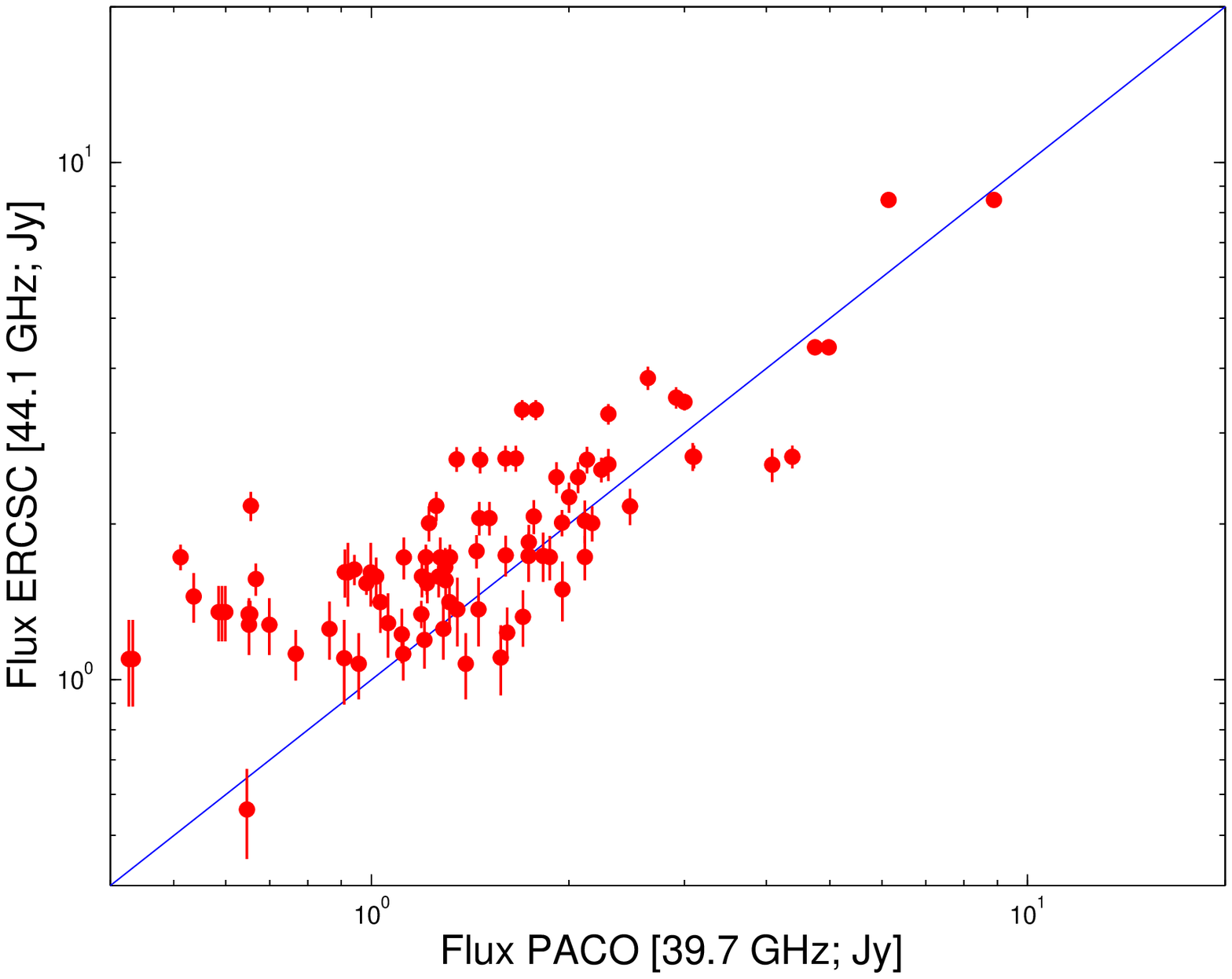}
\caption{Comparison between the ERCSC flux densities at 30\,GHz (left
panel) and at 44\,GHz (right panel) with the almost simultaneous ATCA
measurements (PACO project) at 32.2 and 39.7\,GHz, respectively. No
correction for the slightly different frequencies has been applied.
  \label{fig:paco}}
\end{center}
\end{figure*}

The ``flat'' spectra of blazars are generally believed to result
from the superposition of different components in the inner part of
AGN relativistic jets, each with a different synchrotron
self-absorption frequency \citep{KellermannPauliny-Toth69}. At a
given frequency, the observed flux density is thus dominated by the
synchrotron-emitting component which becomes self-absorbed and, in
the equipartition regime, the resulting spectrum is approximately
flat. However, this ``flat'' spectrum cannot be maintained up to
very high frequencies, because of electron energy losses in the
dominant jet-emission component (i.e., electron ageing), or the
transition to the optically-thin regime, with the onset of a
``steep'' spectrum with a standard spectral index $\alpha= -0.7$ to
$-0.8$. A slightly steepened spectrum may also be caused by the
superposition of many jet components. The redshift moves the
observed steepening to lower frequencies and, thus, a greater
fraction of blazar sources are observed with a steep spectrum at
sub-mm wavelengths. With current data it is not yet possible to
decide among the different scenarios. However, given their
sensitivity and full sky coverage, \Planck\ surveys are uniquely
able to shed light on this transition from an almost ``flat'' to a
``steep'' regime in the spectra of blazar sources.

The outline of this paper is as follows. In \S\,\ref{sec:over} we
briefly sketch the main properties of the ERCSC. In
\S\,\ref{sec:valid} we summarize the source validation. In
\S\,\ref{sec:sample} we describe the complete sample selected at
30\,GHz, used for the analysis of spectral properties. In
\S\,\ref{sec:counts} we present the source counts over the frequency
range 30--217\,GHz.  In \S\,\ref{sec:spectra}  we investigate the
spectral index distributions in different frequency intervals.
Finally, in \S\,\ref{sec:concl} we summarize our main conclusions.

\section{The \Planck\ ERCSC}\label{sec:ercsc}

\subsection{Overview}\label{sec:over}

The \Planck\ ERCSC \citep{planck2011-1.10} lists positions and flux
densities for the compact sources recovered from the \Planck\ first
1.6 full sky survey maps in nine frequency channels between 30 and
857\,GHz. Thus about 60\% of the sources have been covered twice,
with a time separation of about 6 months. Sources near the ecliptic
poles, where the scan circles intersect, are often covered multiple
times. ERCSC flux densities are therefore averages over different
observing time periods. They have been calculated by aperture
photometry, using the most recent definition of the beam shapes and
sizes \citep{planck2011-1.7, planck2011-1.6}. In the
frequency range considered in this paper, the \Planck\ photometric
calibration is based on the CMB dipole and on the modulation induced
on it by the spacecraft orbital motion. According to
\citet{planck2011-1.6} the absolute photometric calibration at LFI
frequencies ($30$, $44$ and $70$ GHz) is at the $\sim 1\%$ level,
while \citet{planck2011-1.7} reports on a relative photometric
accuracy, between the frequency channels from 100 to 353 GHz, better
than $2\%$ and, more likely, at the $\sim 1\%$ level.

The final version of the ERCSC (see also \citealt{planck2011-1.10}
for more details) has been created by specifically applying the
so-called ``PowellSnakes''detection method \citep{Carvalho09} to the
\Planck\ full sky (dipole subtracted) anisotropy maps in the
frequency channels from 30 to 143\,GHz and the SExtractor package
\citep{Bertin96} in the four channels at higher frequencies (from
217 to 857\,GHz). A ``compact'' source is accepted in the ERCSC if
it survived a set of primary and secondary selection criteria. The
primary criterion utilized the feedback from the Monte Carlo Quality
Assessment system and introduced a signal-to-noise ratio cut to
ensure that $>90\%$ of the sources in the catalogue are reliable and
have a flux density accuracy better than 30$\%$. The secondary
criterion comprised a set of cuts that removed the extended sources
with an elongation $>3.0$ (the ratio between the major and minor
axis, in pixels, of the detected source) and sources that could be
potentially spurious. For instance a source is dropped whenever
$>5\%$ of the pixels within 2$\times$FWHM of its position had
invalid values (see the ERCSC Explanatory Supplement for more
details). The overriding requirement in constructing the ERCSC is
source reliability and not completeness.

\subsection{Validation and photometry check of ERCSC sources}\label{sec:valid}

\begin{figure*}
\begin{center}
\includegraphics[width=0.95\textwidth]{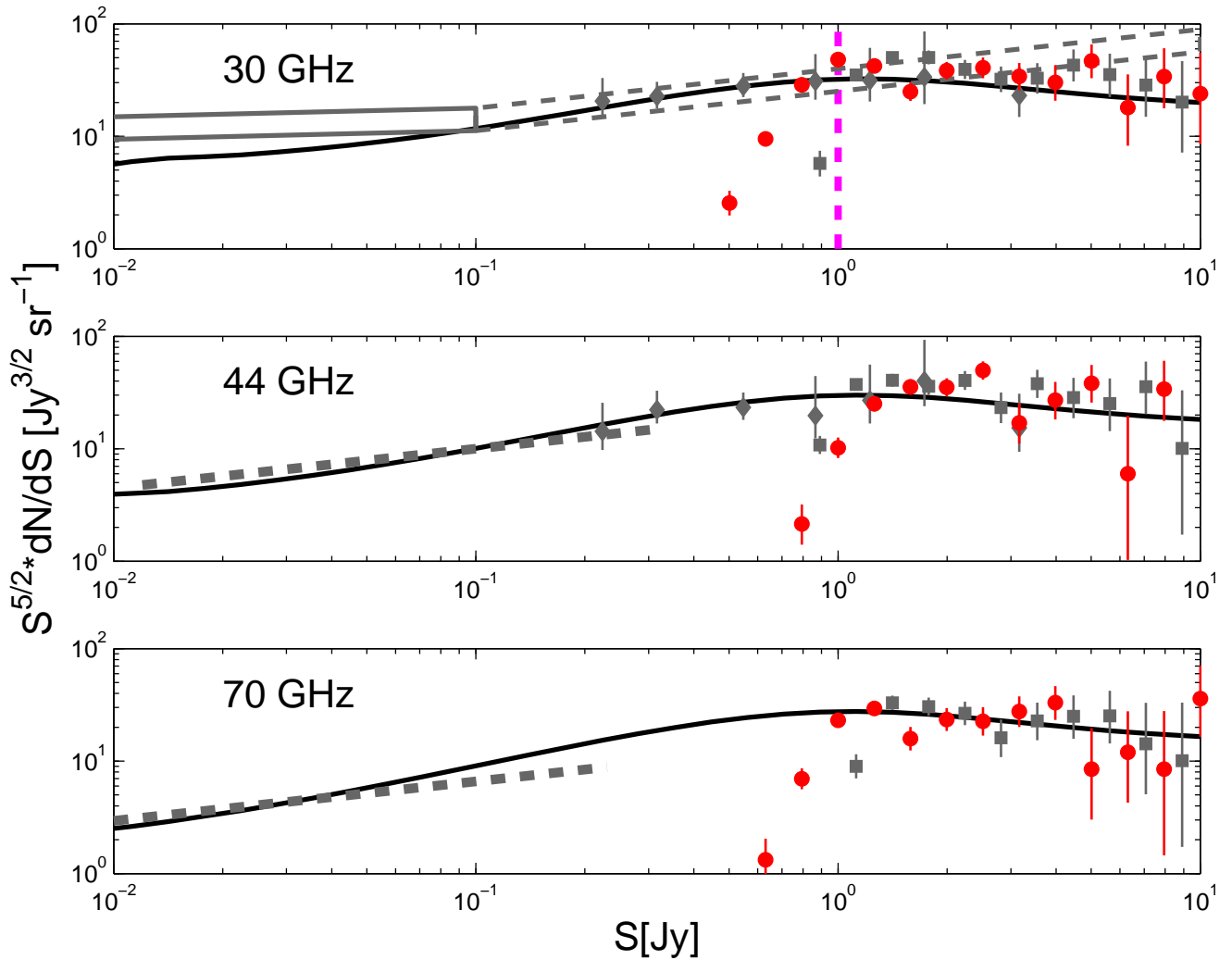}
\caption{Euclidean normalized differential number counts at the LFI
frequencies. The red circles with Poisson error bars show the counts
of sources with counterparts in our reference 30\,GHz sample. In
each panel, the solid curves show the total number counts of
extragalactic radio sources predicted by the \cite{DeZotti05} model.
Also shown are: the counts estimated at 31\,GHz from DASI (grey
dashed box; \cite{Kovac02}) and at 33\,GHz from the VSA data (grey
box; \cite{Cleary05}); the counts from the PACO (grey diamonds;
Bonavera et al., in preparation) and the \textit{WMAP} 5-yr surveys
(grey squares; \cite{Massardi09}), at the closest frequencies, and
the counts estimated by \cite{Waldram07} (grey dashed line),
exploiting multi-frequency follow-up observations of the 15
GHz 9C sources. The vertical dashed magenta line in the upper panel
indicates the flux density completeness limit, 1.0\,Jy, estimated
for our primary sample \S\,\ref{sec:sample}.
  \label{fig:counts_low}}
\end{center}
\end{figure*}

The validation process for the compact sources included in the
\Planck\ ERCSC was performed by two different teams, selected among
members of the \Planck\ Consortia: a Validation Team on radio
sources (VTRS) and a Validation Team on far-IR sources. The two
teams worked separately at first, but cross-checked their results in
the second phase of the process. The processing steps and main
outcomes are summarized in the Explanatory Supplement released with
the ERCSC. For compact radio (i.e., synchrotron dominated) sources,
the VTRS (see \cite{planck2011-6.2} for a more detailed discussion)
has found that $> 97\%$
of the ERCSC sources at 30\,GHz have reliable counterparts in published
catalogues at GHz frequencies (PMN: \citealt{Wright96}; GB6:
\citealt{Gregory96}; NVSS: \citealt{Condon98}; SUMSS: \citealt{Mauch03};
AT20G: \citealt{Massardi08,Murphy10}). Similar (although slightly lower)
percentages were found for ERCSC sources detected at 44 and 70\,GHz.
At higher frequencies ($\ge 100$\,GHz) \Planck\ detects an increasing fraction
of dusty galaxies, undetected by low-frequency surveys. Therefore, the
source reliability was confirmed by internal
matches of sources detected in two neighboring \Planck\ frequency
channels: i.e., 143 and 217\,GHz, or 217 and 353\,GHz, etc.. However, the
validation of synchrotron-dominated sources is still relatively easy to perform,
since all of them must be present in low-frequency catalogues.

The \textit{WMAP} 7-year catalogue \citep{gold2010} contains a total
of 471 sources detected in at least one frequency channel. Of these,
289, 281, 166 and 59 sources are detected as $\geq 5\sigma$ peaks in
the 33, 41, 61, and 94\,GHz maps, respectively. The ERCSC catalogue
includes 88$\%$, 63$\%$, 81$\%$, and 95$\%$ of the $5\sigma$
\textit{WMAP} sources at 30, 44, 70, and 100 GHz,respectively. The
median of the distribution of offsets between \textit{WMAP}
and \Planck\ positions at each frequency are $2.5'$, $2.1'$, $1.7'$,
and $1.0'$ at each frequency (see also \citealt{planck2011-6.2} for
a more detailed discussion on this subject). Except for the 44\,GHz
channel, where \Planck\ is known to be less sensitive, most
\textit{WMAP} sources that failed to be included in the \Planck\
ERCSC (31 sources at 30\,GHz) are generally at the faint end of the
flux density distribution (i.e., near the detection threshold) and
may have flux densities boosted by the Eddington bias or the effects
of confusion, or may be spurious. The absence from the ERCSC of a
few   brighter \textit{WMAP} sources (5 sources at 30\,GHz) is
probably caused by their variability.

The \Planck-ATCA Co-eval Observations (PACO) project
\citep{Massardi11} has provided measurements with the Australia
Telescope Compact Array (ATCA) of sources potentially detectable by
\Planck\ almost simultaneously with \Planck\ observations. 147 ERCSC
sources have PACO observations at 32.2 and 39.7\,GHz within 10 days
of \Planck\ observations. All these sources are unresolved also by
ATCA. As illustrated in Fig. \ref{fig:paco}, the comparison between
ATCA and ERCSC flux densities at the nearest frequencies (30 and 44
\,GHz, respectively), shows a reassuringly close agreement. The faintest ERCSC flux densities are obviously enhanced by the effect of the Eddington bias,the noise-increased number count of point sources to a given detection threshold, which is besides enhanced in the 44 GHz \Planck\ channel, where the noise level is higher.

\subsection{The 30\,GHz extragalactic radio source sample}\label{sec:sample}
\subsubsection{Identification of compact Galactic sources}

To minimize the contamination of the sample by Galactic sources we
have restricted ourselves to $|b|> 5^\circ$ and we have also
excluded sources within $5^\circ$ and $2.5^\circ$, respectively,
of the nominal centres of the Large and Small Magellanic Clouds.
Outside these regions, a search in the SIMBAD database, with a
search radius of $16'$, corresponding to about half the FWHM at 30\,GHz,
has yielded 18 associations of ERCSC sources with known Galactic
objects (5 PNe, 10 HII regions, and 3 SNRs), all within $5'$ of the
ERCSC position. After having removed these sources we are left with
533 compact extragalactic radio sources detected at 30\,GHz, with
$97\%$ or more of them identified in external catalogues at GHz
frequencies. This constitutes our primary sample.

\subsubsection{Completeness and uniformity tests}

\begin{figure}
\begin{center}
\includegraphics[width=0.5\textwidth]{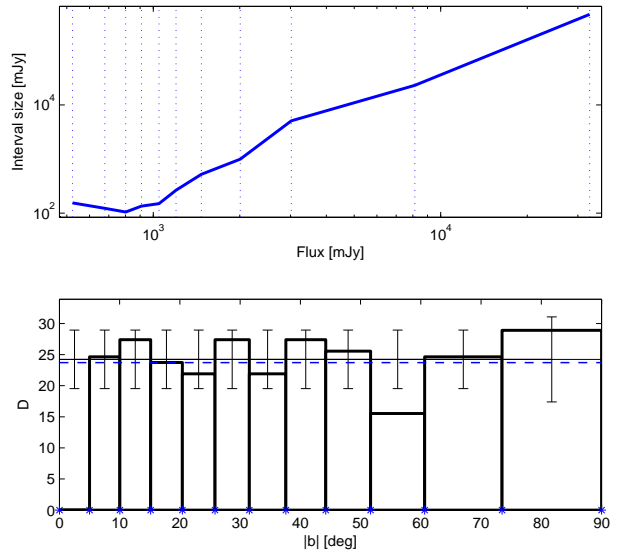}
\caption{Completeness and uniformity tests at 30\,GHz. Top panel:
flux density interval, or bin size, including a fixed number of
sources as a function of flux density.  The change of slope at about
0.9\,Jy signals the onset of incompleteness. Bottom panel:
uniformity test for sources with flux densities $S_\nu\geq 1.0$\,Jy.
The source density, $D$, in sources per sr, within regions at
different Galactic latitudes, shows an acceptable uniformity.
The horizontal grey (solid) and blue (dashed) lines show the
average source density at Galactic latitudes above $\vert b\vert =5$
and $\vert b\vert =30$ deg, respectively. As it is apparent, the two
values are very close to each other, within the $1\sigma$ level calculated for the overall population in the two sky areas considered here, and well inside the $1\sigma$
normalized Poisson error bars. We also checked that at flux
densities below about 0.9--1.0\,Jy statistically more relevant
deviations from a uniform distribution begin to appear.
\label{fig:compl}}
\end{center}
\end{figure}

An indication of the completeness limit of our sample is obtained by
looking at the differential counts (see top panel in Fig.
\ref{fig:counts_low}) : a sharp decrease of the slope at faint flux
densities ($S_\nu\lsim 1$\,Jy at 30\,GHz) signals the onset of
incompleteness. Based on the ATCA 20\,GHz counts
\citep{Massardi08,Murphy10}, we expect that the slope of the counts
remains approximately constant over the limited flux density range
covered by the ERCSC. Therefore, the flux density interval (or bin
size) containing a fixed number of sources must decrease as a power
of the central flux density. As shown in the top panel of Fig.
\ref{fig:compl}, at 30\,GHz this happens down to a flux density of
about 0.9\,Jy, where the curve abruptly flattens.

We need also to test whether the spatial distribution of sources in
our sample is consistent with being statistically uniform, as it
must be in the case of extragalactic sources. Deviations from
uniformity may be expected at lower Galactic latitudes, both because
of residual contamination by unrecognized Galactic sources and
through the effect of a stronger Eddington bias due to fluctuations
of diffuse Galactic emission. The bottom  panel of Fig.
\ref{fig:compl} shows no significant deviations from a uniform
distribution on the sky for $|b|> 5^\circ$ if we adopt a
completeness limit of $\simeq 1.0$\,Jy at 30\,GHz.
Remarkably, the average source density at $\vert b\vert
>$ 5$^\circ$, $D=24.23$ (in sources per sr), is very similar to the value
found at $\vert b\vert>$ 30$^\circ$, $D=23.71$, which guarantees
that we are not losing extragalactic sources, in this frequency
range, when going down to lower Galactic latitudes, and that the
residual contamination due to unrecognized Galactic sources is
negligible. Larger deviations from uniformity are found if we adopt
fainter completeness limits. Taking into account both results, we
therefore adopt $S_\nu$=1.0 Jy as an estimate for the completeness
limit at 30\,GHz. Our primary sample is made of 290 sources above
the adopted $S_\nu=1.0$\,Jy flux density limit. As a comparison,
\citet{Massardi09} detected 281 sources at $\vert b\vert>5^{\deg}$
and $S_\nu\gsim 1$\,Jy in their blind survey performed on the 5-year
\textit{WMAP} 33\,GHz map.

\begin{figure}
\begin{center}
\includegraphics[width=0.45\textwidth]{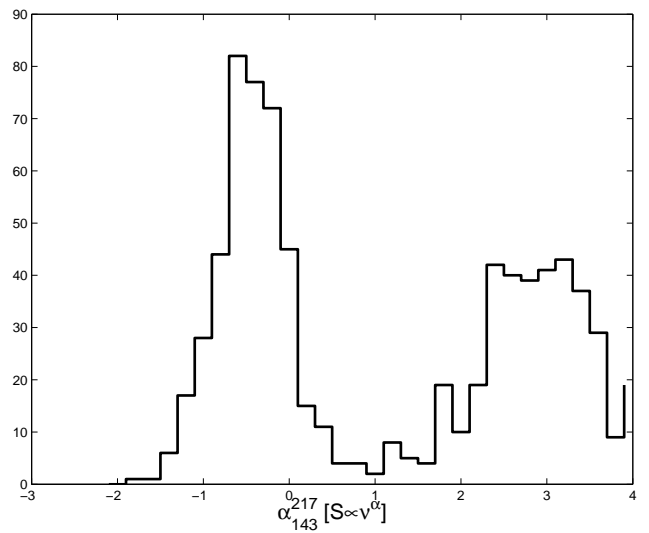}
\caption{Spectral index distribution of ERCSC sources between 143
and 217GHz. Only sources detected at 143\,GHz and at 217\,GHz have
been considered: no upper limits on flux densities have been used in
this calculation.
  \label{fig:idx_ir}}
\end{center}
\end{figure}

In addition, we used the \Planck\ HFI frequency channels at 143 and
217\,GHz to select all the extragalactic sources in the ERCSC whose
spectra are still dominated by non-thermal synchrotron emission at
217\,GHz (409 sources with $\alpha_{143}^{217}<0.5$; see Fig.
\ref{fig:idx_ir}). We limit the selection of this secondary sample
to the above $\alpha_{143}^{217}$ value with the purpose of
excluding all possible sample contamination coming from a second
population of sources dominated by thermal dust emission
($\alpha_{143}^{217}>1.0$). The choice of $\alpha_{143}^{217}<1.0$,
corresponding to a minimum in the distribution, would not change our
results since very few sources ($< 10$) have spectral indices in the
interval $0.5<\alpha_{143}^{217}<1.0$. This selection (at above
$100$\,GHz, where there is not yet a complete match of ERCSC sources
with external catalogs) is useful for comparing the outcomes from
this (secondary) sample with our predictions on the
statistical properties of extragalactic sources in our primary
sample, selected at\,30\,GHz.

\section{Number counts}\label{sec:counts}

\begin{table*}
\begingroup
\newdimen\tblskip \tblskip=5pt
\caption{Euclidean-normalized differential number counts per
steradian estimated from the \Planck\ ERCSC at 30--217\,GHz. The 143
and 217 GHz number counts are those represented by blue diamonds in
Fig. 5. The bins are centred in the $\log_{10}($S$_\nu)$ values and are
symmetric in logarithmic scale.}
\label{tab:counts} 
\nointerlineskip \vskip -3mm \footnotesize
\setbox\tablebox=\vbox{
\newdimen\digitwidth
\setbox0=\hbox{\rm 0}
\digitwidth=\wd0
\catcode`*=\active
\def*{\kern\digitwidth}
\newdimen\signwidth
\setbox0=\hbox{+}
\signwidth=\wd0
\catcode`!=\active
\def!{\kern\signwidth}
\halign{\hfil#\hfil\tabskip=2em&\hfil#\hfil&\hfil#\hfil&\hfil#\hfil&\hfil#\hfil&\hfil#\hfil &\hfil#\hfil\tabskip=0pt\cr
\noalign{\doubleline} &\multispan6\hfil $S^{5/2}dN/dS\,[$Jy$^{3/2}
$sr$^{-1}]$ \hfil\cr \noalign{\vskip -3pt} &
\multispan6\hrulefill\cr \noalign{\vskip 2pt} $\log_{10}$(S$_\nu$) &
30 & 44 & 70 & 100 & 143 & 217 \cr [Jy] & [GHz] & [GHz] & [GHz] &
[GHz] & [GHz] & [GHz] \cr \noalign{\vskip 3pt\hrule\vskip 5pt}
$-0.5$ & ... & ... & ... & ... & *$ 0.9 \pm  0.2$ & *$ 4.6 \pm 0.6$
\cr

$-0.4$ & ... & ... & ... & *$ 1.1 \pm  0.4$ & *$ 6.3 \pm  0.8$ & *$
8.2 \pm  0.9$ \cr

$-0.3$ & $ 2.6 \pm  0.6$ & ... & ... & *$ 5.8 \pm  0.9$ & $11.6 \pm
1.2$ & *$ 8.9 \pm  1.1$ \cr

$-0.2$ & $ 9.5 \pm  1.3$ & ... & $ 1.3 \pm  0.6$ & $13.3 \pm  1.6$ &
$ 11.6 \pm  1.5$ & *$ 8.5 \pm  1.3$ \cr

$-0.1$ & $29 \pm  *3$ & $ 2.1 \pm  0.7$ & $ 7.0 \pm  1.4$ & *$20 \pm
*2$ & *$12.6 \pm  1.8$ & $10.5 \pm  1.7$ \cr

!$ 0.0$ & $48 \pm *4$ & $10 \pm *2$ & $23 \pm *3$ & *$25 \pm *3$ &
*$16 \pm *2$ & *$11 \pm *2$ \cr

!$ 0.1$ & $42 \pm *5$ & $25 \pm *4$ & $29 \pm *4$ & *$22 \pm *3$ &
*$12 \pm *2$ & *$14 \pm *3$ \cr

!$ 0.2$ & $25 \pm *4$ & $36 \pm *5$ & $16 \pm *3$ & *$18 \pm *4$ &
*$15 \pm *3$ & *$*7 \pm *2$ \cr

!$ 0.3$ & $38 \pm *6$ & $35 \pm *6$ & $24 \pm *5$ & *$18 \pm *4$ &
*$20 \pm *5$ & *$12 \pm *4$ \cr

!$ 0.4$ & $41 \pm *8$ & $50 \pm *9$ & $23 \pm *6$ & *$21 \pm *6$ &
*$11 \pm *4$ & *$*6 \pm *3$ \cr

!$ 0.5$ & $34 \pm *8$ & $17 \pm *6$ & $28 \pm *8$ & *$15 \pm *6$ &
*$11 \pm *5$ & *$13 \pm *5$ \cr

!$ 0.6$ & $30 \pm *9$ & $27 \pm *9$ & $33 \pm 10$ & *$21 \pm *9$ &
*$12 \pm *6$ & *$*9 \pm *5$ \cr

!$ 0.7$ & $47 \pm 14$ & $38 \pm 12$ & $ *9 \pm *6$ & *$ *4 \pm *4$ &
*$13 \pm *7$ & *$13 \pm *7$ \cr

!$ 0.8$ & $18 \pm 10$ & $ *6 \pm *5$ & $12 \pm *8$ & *$18 \pm 10$ &
*$12 \pm *8$ & *$12 \pm *8$ \cr

!$ 0.9$ & $34 \pm 16$ & $34 \pm 16$ & $ *8 \pm *7$ & *$17 \pm 11$ &
*$*8 \pm *7$ & *$*8 \pm *7$ \cr

!$ 1.0$ & $24 \pm 15$ & ... & $36 \pm 20$ & *$36 \pm 20$ & *$24 \pm
15$ & ... \cr \noalign{\vskip 5pt\hrule\vskip 3pt}}}
\endPlancktable 
\endgroup
\end{table*}

\begin{figure*}
\begin{center}
\includegraphics[width=0.95\textwidth]{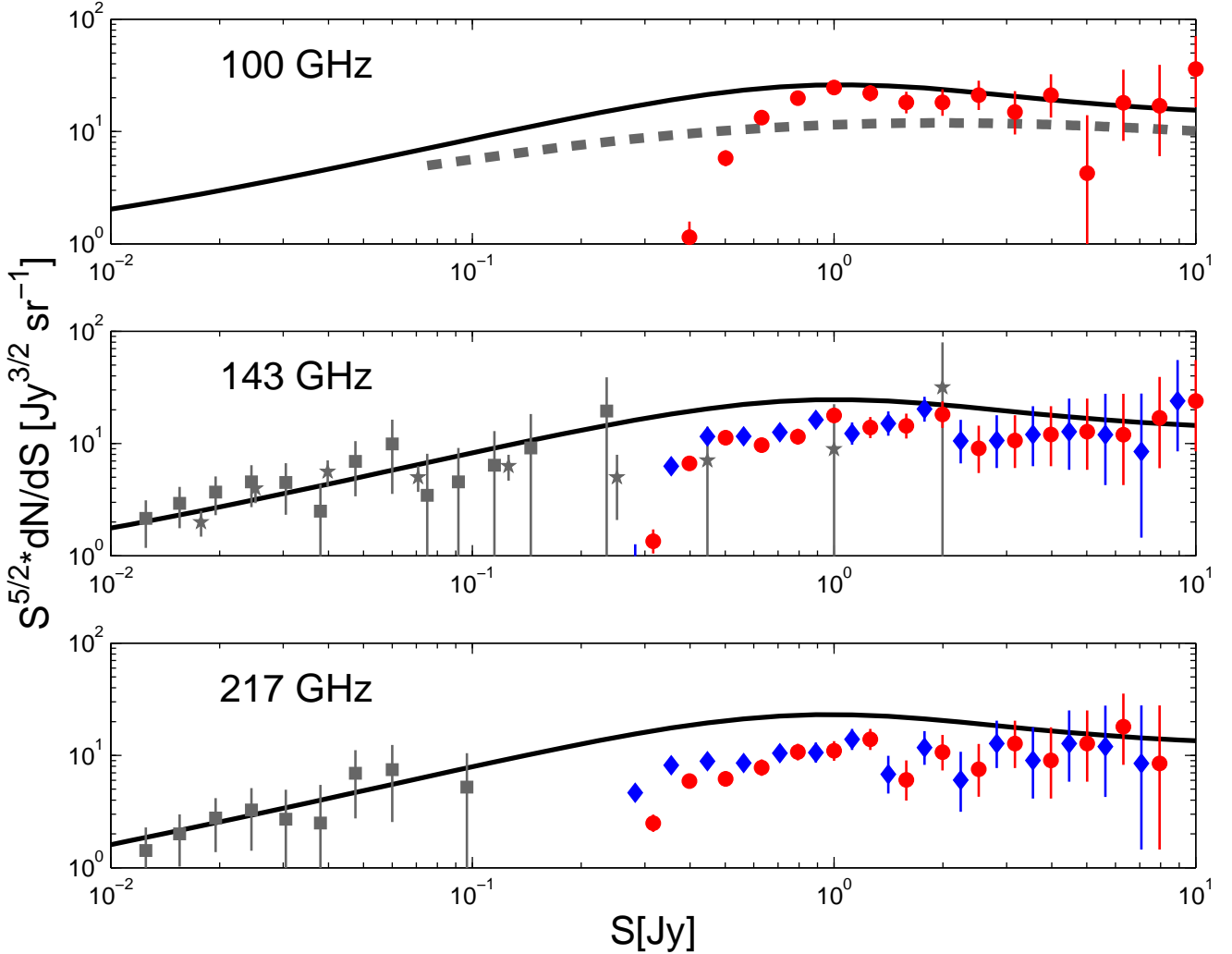}
\caption{Euclidean normalized differential number counts at the HFI
frequencies (100, 143, and 217\,GHz). The red circles with Poisson
error bars show the counts of sources with counterparts in our
reference 30\,GHz sample. At 143 and 217\,GHz the blue diamonds
(shifted to the left by half of the bin size, for clarity)
show the counts obtained after removing sources with 143--217\,GHz
spectral index indicative of dust emission (see Sect. 2.3.2). Again,
in each panel, the solid curves show the total number counts of
extragalactic radio sources predicted by the \cite{DeZotti05}
evolution model. Also shown are the SPT (grey squares;
\citealt{Vieira10}) and ACT (grey stars; \citealt{Marriage10} counts
of radio sources. At 100\,GHz we also show the estimated counts by
\cite{Sadler08} from follow-up observations of a sample of
sources selected from the 20\,GHz ATCA survey (grey dashed line).
  \label{fig:counts_high}}
\end{center}
\end{figure*}

Figures~\ref{fig:counts_low} and \ref{fig:counts_high} show the
number counts of extragalactic radio sources at the six \Planck\
frequencies from 30 to 217\,GHz (see also Table \ref{tab:counts}).
The sharp breaks in the number counts at approximately 1.0\,Jy
(30\,GHz), 1.5\,Jy (44\,GHz), 1.1\,Jy (70\,GHz), 0.9\,Jy (100\,GHz),
0.5\,Jy (143\,GHz) and 0.4\,Jy (217\,GHz) signal the onset of
incompleteness.

The results of deeper surveys on small fractions of the sky and the
\textit{WMAP} differential number counts are also shown, for
comparison. The agreement with \textit{WMAP} data is very good. Our
differential counts at 30 and 70\,GHz are somewhat deeper than the
\textit{WMAP} ones at 33 and 61\,GHz, while at 44\,GHz they are
somewhat shallower than the ones calculated for the 41\,GHz
\textit{WMAP} channel. Also our counts above the completeness limits
appear to join smoothly with those from deeper surveys.

At frequencies of up to 100\,GHz, the predictions of the
\cite{DeZotti05} cosmological evolution model -- relying on
extrapolations from lower frequency data and capable of providing a
very good fit to almost all data on number counts as well as on
other statistics of radio sources at frequencies above 5\,GHz -- are
in generally good agreement with our current findings. This result
implies that no new radio source population shows up at bright flux densities.
Very few ``extreme'' or ``inverted--spectrum'' compact radio sources
are found in the \Planck $\ $ERCSC. The emission
and spectral properties of these sources, which are interesting in their own right, are
discussed in a companion paper \citep{planck2011-6.2}.

At higher frequencies (i.e., at 143 and 217\,GHz) we also
plotted the number counts obtained by using the sample of radio
sources selected at 143 and 217 GHz (blue diamonds; see
\S\,\ref{sec:sample} and Fig. \ref{fig:idx_ir} for more details).
These number counts (calculated from our secondary sample, 284 sources at above $0.5$ Jy) turn out
to be in almost perfect agreement with the ones obtained in the previous
Section from our primary sample selected at 30\,GHz (290 sources at $S_\nu\geq1.0$), thus confirming
that no bright extragalactic radio sources are missed by our
selection criteria and that the underlying parent population turns
out to be (statistically) the same.

Figure \ref{fig:counts_high} shows that the \cite{DeZotti05} model
over-predicts the bright counts by a factor of about 2 at 143\,GHz
and about 2.6 at 217\,GHz, while it is consistent with the fainter
SPT \citep{Vieira10} and ACT \citep{Marriage10} counts. As discussed
in the next Section, the discrepancy between the model and our
current data is due to a steepening of the spectra of ERCSC sources
above about 70\,GHz, not predicted by the model but, at least
partially, already suggested by other data sets
\citep{GonzalezNuevo08,Sadler08}.

An implication of this result is that the contamination of the CMB
angular power spectrum by extragalactic radio sources below the
detection limit at 143 and 217\,GHz is lower than predicted by the
\cite{DeZotti05} model. Assuming a Poisson distribution (clustering
effects are reduced to negligible values by the very broad
luminosity function of radio sources, e.g.,
\cite{Toffolatti05,DeZotti10}) and simply scaling down the model
counts by the factors mentioned above, the amplitude of the angular
power spectrum of unresolved sources goes down by roughly
the same factor \footnote{As an example, the amplitude of the power
spectrum of radio sources, in terms of $C_\ell$ values, below, e.g.,
$S_\nu\leq1$\,Jy at 217\,GHz, corresponds to $\sim 4.2\times 10^{-5}
\mu$K$^2$, if calculated from the \Planck\ ERCSC differential number
counts of Table \ref{tab:counts}. For comparison, undetected radio
sources sum up $\sim 1.1\times 10^{-4} \mu$K$^2$, if we integrate
the differential number counts of the \cite{DeZotti05} cosmological
evolution model up to $S_\nu=1$\,Jy.}. This is, however, an upper
limit to the correction factor, especially at 217\,GHz, because if
we apply the factor calculated above to all flux densities we would
end up with a clear underestimate of the faint counts measured by
\cite{Vieira10}. Therefore, if we consider much fainter source
detection limits, as foreseen for future experiments in the sub-mm,
the amplitude of the angular power spectrum due to unresolved
sources stays essentially at the same level as predicted by the De
Zotti et al. model.

\section{Spectral index distributions}\label{sec:spectra}

\begin{table} 
\begingroup 
\newdimen\tblskip \tblskip=5pt
\caption{Median and standard deviations of the spectral index
distributions between 30\,GHz and the selected frequency. We adopt
the convention $S_\nu\propto \nu^\alpha$.}
\label{tab:spec_index_evo}
\nointerlineskip
\vskip -3mm
\footnotesize 
\setbox\tablebox=\vbox{ %
\newdimen\digitwidth 
\setbox0=\hbox{\rm 0}
\digitwidth=\wd0
\catcode`*=\active
\def*{\kern\digitwidth}
\newdimen\signwidth
\setbox0=\hbox{+}
\signwidth=\wd0
\catcode`!=\active
\def!{\kern\signwidth}
\halign{\bf\hfil#\hfil\tabskip=1.5em&\hfil#\hfil&\hfil#\hfil&\hfil#\hfil&\hfil#\hfil&\hfil#\hfil\tabskip=0pt \cr 
\noalign{\doubleline}
$\nu$[GHz] & *44 & *70 & *100 & *143 & *217 \cr
\noalign{\vskip 3pt\hrule\vskip 5pt}
median & $-0.06$ & $-0.18$ & $-0.28$ & $-0.39$ & $-0.37$ \cr
error & !0.01 &  !0.01 &  !0.01 &  !0.01 &  !0.01 \cr
$\sigma$ & !0.30 & !0.18 & !0.17 & !0.16 & !0.15 \cr
\noalign{\vskip 5pt\hrule\vskip 3pt}}}
\endPlancktable 
\endgroup
\end{table}

To study the spectral properties of the extragalactic radio sources
in the \Planck\ ERCSC we used our reference 30\,GHz sample above the
estimated completeness limit ($1.0\,$Jy; \S\,\ref{sec:sample}). Not
all of these sources were detected at the $\ge 5\sigma$ level in
each of the \Planck\ frequency channels considered. Whenever a
source was not detected in a given channel we replaced its (unknown)
flux density by a $5\sigma$ upper limit, where for $\sigma$ we used
the average r.m.s. error estimated at each \Planck\ frequency. The
upper limits have been redistributed among the flux density bins by
using a Survival Analysis technique and, more specifically, by
adopting the Kaplan-Meyer estimator (as implemented in the ASURV
code, \citep{Lavalley92}). Since the fraction of upper limits is
always small (it reaches approximately $30\%$ only in our less
sensitive channel at 44GHz), the spectral index distributions are
reliably reconstructed at each frequency.

Table~\ref{tab:spec_index_evo} gives the median spectral indices
between 30\,GHz and the other frequencies considered here and the
dispersions of the spectral index distributions. A moderate
steepening of spectral indices at higher frequencies is clearly
apparent. Hints in this direction were previously found by
\cite{GonzalezNuevo08} from their analysis of the NEWPS sample
\citep{LopezCaniego07} and also by \cite{Sadler08}. Additional
evidence of spectral steepening is presented in \cite{planck2011-6.3a}.

Our 30--100\,GHz spectral index is close to the $\alpha\simeq -0.39$
found by \cite{Sadler08} between 20 and 95\,GHz, for a sample with
20\,GHz flux density $S>150\,$mJy. Moreover, our 30--143\,GHz median
spectral index is in very good agreement with the one found by
\cite{Marriage10} for their bright ($S_\nu>50$ mJy)
148\,GHz-selected sample with complete cross-identifications from
the Australia Telescope 20\,GHz survey, i.e
$\alpha_{20}^{148}=-0.39\pm 0.04$. On the other hand,
\citep{Vieira10} find that their much fainter synchrotron emitting
radio sources selected at 150\,GHz are consistent with a flatter
spectral behaviour (mean $\alpha_5^{150}\simeq -0.1$) between 5\,GHz
and 150\,GHz.  \cite{Massardi10} find mean spectral indices
$\alpha_5^{150}\simeq -0.17$ and $\alpha_{20}^{150}\simeq -0.30$ for
AT20G sources with 150\,GHz flux density $S>50\,$mJy. A flattening
of the mean/median high-frequency spectral indices at flux densities
fainter than the ones probed by the \Planck\ ERCSC may help to
account for the unusually ``flat'' normalized counts at 143 and
217\,GHz.

\begin{figure}
\begin{center}
\includegraphics[width=0.45\textwidth]{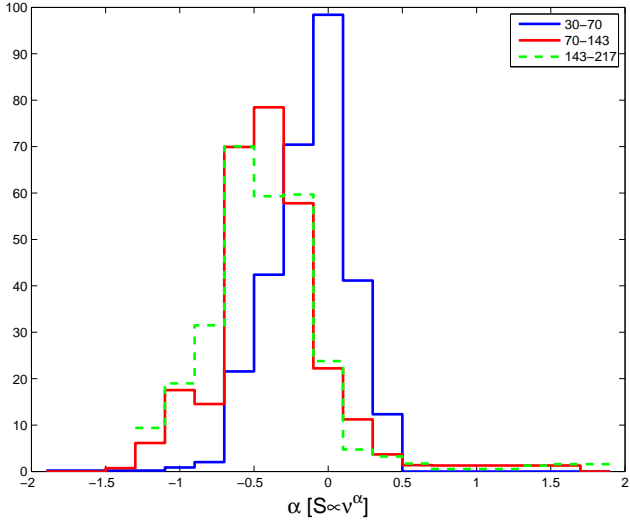}
\caption{Spectral index distributions for different frequency
intervals calculated by taking into account all sources selected at
30\,GHz with $S_\nu> 1$\,Jy. There is clear evidence for a
steepening above 70\,GHz.
  \label{fig:spec_idx}}
\end{center}
\end{figure}

In Fig.~\ref{fig:spec_idx} we compare the distributions of  spectral
indices over different frequency intervals. There is a clear shift
toward steeper values above 70\,GHz: the median values vary from
$\alpha_{30}^{70}= -0.18 \pm 0.01$ ($\sigma= 0.18$) to
$\alpha_{70}^{143}= -0.52 \pm 0.01$ ($\sigma = 0.22$). On the other
hand, the distribution of spectral indices between 143 and 217\,GHz
is close to the one found for $\alpha_{70}^{143}$
($\alpha_{143}^{217}= -0.46 \pm 0.01$; $\sigma$= 0.23). This latter
result is again very similar to the corresponding value calculated
for all the sources detected at 143 and 217\,GHz with
$\alpha_{143}^{217}<0.5$, i.e., $\alpha_{143}^{217}= -0.51 \pm
0.01$, as shown in Fig. \ref{fig:idx_ir}. Moreover, in the paper
\cite{planck2011-6.3a} an average value of $\alpha=-0.56$
($\sigma=0.29$) at HFI frequencies is found for their sample of 84
bright blazars selected at 37\,GHz, with flux densities measured in
at least 3 HFI channels, in full agreement with our present
findings.

\begin{figure}
\begin{center}
\includegraphics[width=0.45\textwidth]{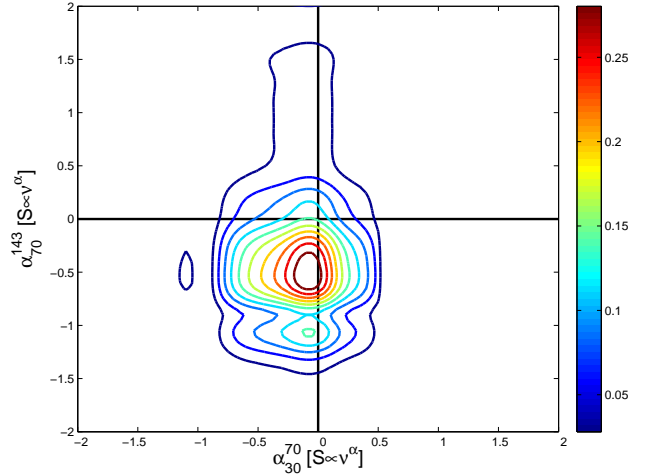}
\caption{Contour levels of the distribution of ${\alpha}_{70}^{143}$ vs.
$\alpha_{30}^{70}$ obtained by Survival Analysis, i.e., taking
into account the upper limits to flux densities at each frequency.
The colour scale can be interpreted as the probability of having any
particular pair of values of the two spectral indices. The maximum
probability corresponds to $\alpha_{30}^{70}\simeq -0.18$ and $\alpha_{70}^{143}\simeq -0.50$.
  \label{fig:colcol}}
\end{center}
\end{figure}

Figure~\ref{fig:colcol} presents the contour levels of the distribution of
$\alpha_{70}^{143}$ vs.  $\alpha_{30}^{70}$  (obtained using
Survival Analysis) in the form of a 2D probability field: the colour
scale can be interpreted as the probability of a given pair of
spectral indices.

As already noted, at high \Planck\ frequencies most of the
extragalactic radiosources are blazars. From the contour plot of
Fig.~\ref{fig:colcol} it is possible to see that the maximum
probability corresponds to $\alpha^{70}_{30}\simeq -0.18$ and
$\alpha^{143}_{70} \simeq -0.5$. A secondary maximum can also be
seen at $\alpha^{143}_{70} \simeq -1.2$. However, a physical
interpretation of these features goes beyond the purposes of this
work and, moreover, more data at higher frequencies are needed. A
detailed discussion on the modelling of the spectra of this source
population is presented in a companion paper
\citep{planck2011-6.3a}.

\section{Conclusions}\label{sec:concl}

We have exploited the ERCSC to estimate the bright counts of
extragalactic radio sources at 6 frequencies (30, 44, 70, 100, 143,
and 217\,GHz) and to investigate the spectral properties of sources
in a complete sample selected at 30\,GHz. The counts at 30, 44, and
70\,GHz are in good agreement with those derived from \textit{WMAP} data at
nearby frequencies. The completeness limit of the ERCSC is somewhat
deeper than that of \textit{WMAP} at 30 and 70\,GHz and somewhat shallower at
44\,GHz. At higher frequencies the ERCSC has allowed us to obtain the
first estimate of the differential number counts at bright flux density levels.
At 30, 143 and 217\,GHz, the present counts join smoothly to those
from deeper surveys over small fractions of the sky.

The \cite{DeZotti05} model is consistent with the present counts at
frequencies up to 100\,GHz, but over-predicts the counts at higher
frequencies by a factor of about 2.0 at 143\,GHz and about 2.6 at
217\,GHz. This implies that the contamination of the CMB power
spectrum by radio sources below the 1\,Jy detection limit is
lower than previously estimated. No significant changes are
found, however, if we consider fainter source detection limits, i.e.,
100\,mJy, given the convergence between predicted and observed
number counts.

The analysis of the spectral index distribution over different
frequency intervals, within the uniquely broad range covered by
\Planck\ in the mm and sub-mm domain, has highlighted an average {\it steepening}
of source spectra above about 70\,GHz. The median values of spectral
indices vary from $\alpha_{30}^{70}= -0.18 \pm 0.01$ ($\sigma=
0.18$) to $\alpha_{70}^{143}= -0.52 \pm 0.01$ ($\sigma = 0.22$).
This steepening accounts for the discrepancy between the \cite{DeZotti05}
model predictions and the observed differential number
counts at HFI frequencies. The current outcome is also in
agreement with the findings of \cite{planck2011-6.3a} on a complete
sample of blazars selected at 37\,GHz. The change detected in the
spectral behaviour of extragalactic radio sources in the \Planck\
ERCSC at frequencies above 70-100\,GHz can be tentatively explained
by electron ageing or by the transition to the optically thin
regime, predicted in current models for radio emission in blazar
sources. However, with present data it is not yet possible to
clarify the situation. In the near future, the data of the \Planck\
Legacy Survey will surely prove very useful in settling this open
issue.

\begin{acknowledgements}
The \Planck\ Collaboration thanks the referee, Ronald Ekers,
for his insightful comments, that helped improving the paper. This
research has made use of the SIMBAD database, operated at CDS,
Strasbourg, France. The Planck Collaboration acknowledges the support of: ESA; CNES and CNRS/INSU-IN2P3-INP (France); ASI, CNR, and INAF (Italy); NASA and DoE (USA); STFC and UKSA (UK); CSIC, MICINN and JA (Spain); Tekes, AoF and CSC (Finland); DLR and MPG (Germany); CSA (Canada); DTU Space (Denmark); SER/SSO (Switzerland); RCN (Norway); SFI (Ireland); FCT/MCTES (Portugal); and DEISA (EU).
A description of the Planck Collaboration and a list of its members can be found at
\url{http://www.rssd.esa.int/index.php?project=PLANCK&page=Planck_Collaboration}
\end{acknowledgements}

\bibliographystyle{aa}

\bibliography{Planck_bib}


\end{document}